\documentclass[prl,twocolumn]{revtex4}
\usepackage[dvips]{graphics,color}
\usepackage{amsfonts}
\usepackage{amssymb}
\usepackage{amstext}
\usepackage{epsfig}
\usepackage{amsmath}
\usepackage{graphicx}

\newcommand{\id}{\openone}

\begin{document}

\author{Martin B. Plenio and Shashank Virmani}
\affiliation{Blackett Laboratory, Imperial College London,
Prince Consort Rd, London SW7 2BW, UK}

\affiliation{Institute for Mathematical Sciences, Imperial College
London, 53 Prince's Gate, Exhibition Rd, London, SW7 2PG, UK}
\title{An introduction to entanglement measures.}
\begin{abstract}
We review the theory of entanglement measures, concentrating
mostly on the finite dimensional two-party case. Topics covered
include: single-copy and asymptotic entanglement manipulation; the
entanglement of formation; the entanglement cost; the distillable
entanglement; the relative entropic measures; the squashed
entanglement; log-negativity; the robustness monotones; the
greatest cross-norm; uniqueness and extremality theorems. Infinite
dimensional systems and multi-party settings will be discussed
briefly.
\end{abstract}

\maketitle

\setcounter{page}{1}
\section{Introduction}
The concept of {\it entanglement} has played a crucial role in the
development of quantum physics. In the early days entanglement was
mainly perceived as the qualitative feature of quantum theory that
most strikingly distinguishes it from our classical intuition. The
subsequent development of Bell's inequalities has made this
distinction quantitative, and therefore rendered the non-local
features of quantum theory accessible to experimental verification
\cite{Bell 01,Bell 64,Hardy 98}. Bell's inequalities may indeed be
viewed as an early attempt to quantify the quantum correlations
that are responsible for the counterintuitive features of quantum
mechanically entangled states. At the time it was almost
unimaginable that such quantum correlations could be created in
well controlled environments between distinct quantum systems.
However, the technological progress of the last few decades means
that we are now able to coherently prepare, manipulate, and
measure individual quantum systems, as well as create controllable
quantum correlations. In parallel with these developments, quantum
correlations have come to be recognized as a novel resource that
may be used to perform tasks that are either impossible or very
inefficient in the classical realm. These developments have
provided the seed for the development of modern quantum
information science.

Given the new found status of entanglement as a resource it is quite
natural and important to discover the mathematical structures
underlying its theoretical description. We will see that such a
description aims to provide answers to three questions about
entanglement, namely (1) its characterisation, (2) its manipulation
and, (3) its quantification.

In the following we aim to provide a tutorial overview summarizing
results that have been obtained in connection with these three
questions. We will place particular emphasis on developments
concerning the {\it quantification} of entanglement, which is
essentially the theory of {\it entanglement measures}. We will
discuss the motivation for studying entanglement measures, and
present their implications for the study of quantum information
science. We present the basic principles underlying the theory and
main results including many useful entanglement monotones and
measures as well as explicit useful formulae. We do not, however,
present detailed technical derivations. The majority of our review
will be concerned with entanglement in bipartite systems with finite
and infinite dimensional constituents, for which the most complete
understanding has been obtained so far. The multi-party setting will
be discussed in less detail as our understanding of this area is
still far from satisfactory.

It is our hope that this work will give the reader a good first
impression of the subject, and will enable them to tackle the
extensive literature on this topic. We have endeavoured to be as
comprehensive as possible in both covering known results and also in
providing extensive references. Of course, as in any such work, it
is inevitable that we will have made several oversights in this
process, and so we encourage the interested reader to study various
other interesting review articles (e.g. \cite{Plenio V 98,Schumacher
00,Horodecki 01,Christandl 06,Horodecki H 01,Eisert P 03}) and of
course the original literature.

\medskip

\section{Foundations}
{\bf\em What is entanglement? --} Any study of entanglement
measures must begin with a discussion of what entanglement {\it
is}, and how we actually {\it use} it. In the following we will
adopt a highly operational point of view. Then the usefulness of
entanglement emerges because it allows us to overcome  a
particular constraint that we will call the {\it LOCC constraint}
- a term that we will shortly explain. This restriction has both
technological and fundamental motivations, and arises naturally in
many explicit physical settings involving quantum communication
across a distance.

We will consider these motivations in some detail, starting with the
technological ones. In any quantum communication experiment we would
like to be able to distribute quantum particles across distantly
separated laboratories. Perfect quantum communication is essentially
equivalent to perfect entanglement distribution. If we can transport
a qubit without any decoherence, then any entanglement shared by
that qubit will also be distributed perfectly. Conversely, if we can
distribute entangled states perfectly then with a small amount of
classical communication we may use teleportation \cite{Nielsen C 00}
to perfectly transmit quantum states. However, in any forseeable
experiment involving these processes, the effects of noise will
inevitably impair our ability to send quantum states over long
distances.

\begin{figure}[th]
\centerline{
\includegraphics[width=8cm]{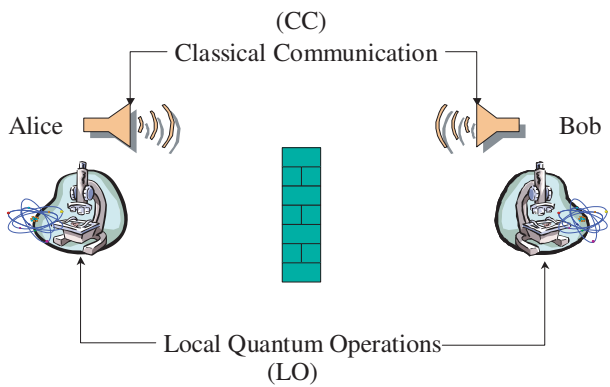}}
\vspace*{0.15cm} \caption{\label{fig2} In a standard quantum
communication setting two parties Alice and Bob may perform any
generalized measurement that is localized to their laboratory and
communicate classically. The brick wall indicates that no quantum
particles may be exchanged coherently between Alice and Bob. This
set of operations is generally referred to as LOCC.}
\end{figure}

One way of trying to overcome this problem is to distribute
quantum states by using the noisy quantum channels that are
available, but then to try and combat the effects of this noise
using higher quality local quantum processes in the distantly
separated labs. Such local quantum operations (`LO') will be much
closer to ideal, as they can be performed in well-controlled
environments without the decoherence induced by communication over
long-distances. However, there is no reason to make the operations
of separated labs totally independent. Classical communication
(`CC') can essentially be performed perfectly using standard
telecom technologies, and so we may also use such communication to
coordinate the quantum actions of the different labs (see fig.
\ref{fig2}). It turns out that the ability to perform classical
communication is vital for many quantum information protocols - a
prominent example being teleportation. These considerations are
the technological reasons for the key status of the {\it Local
Operations and Classical Communication} `LOCC' paradigm, and are a
major motivation for their study.
However, for the purposes of this article, the fundamental
motivations of the LOCC paradigm are perhaps more important than
these technological considerations. We have loosely described
entanglement as the {\it quantum correlations} that can occur in
many-party quantum states. This leads to the question - how do we
define quantum correlations, and what differentiates them from
{\it classical correlations}? The distinction between `quantum'
effects and `classical' effects is frequently a cause of heated
debate. However, in the context of quantum information a precise
way to define classical correlations is via LOCC operations.
Classical correlations can be defined as those that can be
generated by LOCC operations. If we observe a quantum system and
find correlations that cannot be simulated classically, then we
usually attribute them to quantum effects, and hence label them
{\it quantum correlations} \cite{footnote1}. So suppose that we
have a noisy quantum state, and we process it using LOCC
operations. If in this process we obtain a state that can be used
for some task that cannot be simulated by classical correlations,
such as violating a Bell inequality, then we must not attribute
these effects to the LOCC processing that we have performed, but
to quantum correlations that were {\it already present} in the
initial state, even if the initial state was quite noisy. This is
an extremely important point that is at the heart of the study of
entanglement.

It is the constraint to LOCC-operations that elevates entanglement
to the status of a resource. Using LOCC-operations as the only
other tool, the inherent quantum correlations of entanglement are
required to implement general, and therefore nonlocal, quantum
operations on two or more parties \cite{Eisert JPP 00,Collins LP
00}. As LOCC-operations alone are insufficient to achieve these
transformations, we conclude that entanglement may be defined as
the sort of correlations that may not be created by LOCC alone.

Allowing classical communication in the set of LOCC operations means
that they are not completely local, and can actually have quite a
complicated structure. In order to understand this structure more
fully, we must first take a closer look at the notion of general
quantum operations and their formal description.

\begin{figure}[th]
\centerline{
\includegraphics[width=7.5cm]{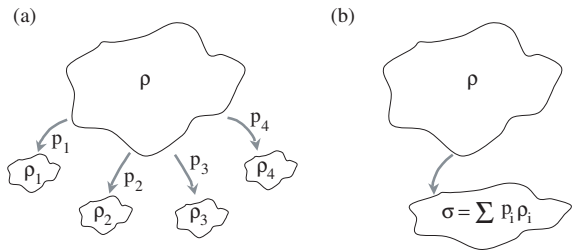}}
\caption{\label{fig1} Schematic picture of the action of quantum
operations with and without sub-selection (eqs. (\ref{eq1}) and
(\ref{eq2}) respectively) shown in part (a) and part (b)
respectively.  }
\end{figure}

{\bf\em Quantum Operations --} In quantum information science much
use is made of so-called `generalised measurements' (see
\cite{Nielsen C 00} for a more detailed account of the following
basic principles). It should be emphasized that such generalised
measurements do not go beyond standard quantum mechanics. In the
usual approach to quantum evolution, a system is evolved according
to unitary operators, or through collapse caused by projective
measurements. However, one may consider a more general setting
where a system evolves through interactions with other quantum
particles in a sequence of three steps: (1) first we first add
ancilla particles, (2) then we perform joint unitaries and
measurements on both the system and ancillae, and finally (3) we
discard some particles on the basis of the measurement outcomes.
If the ancillae used in this process are originally uncorrelated
with the system, then the evolution can be described by so-called
{\it Kraus operators}. If one retains total knowledge of the
outcomes obtained during any measurements, then the state
corresponding to measurement outcomes $i$ occurs with probability
$p_i=tr \{{A_{i}} \rho_{in} {A_{i}}^{\dagger}\}$ and is given by
\begin{equation}
    \label{eq1}
    \rho_{i} = \frac{A_{i} \rho_{in} {A_{i}}^{\dagger}}{tr \{{A_{i}} \rho_{in} {A_{i}}^{\dagger}\}}
\end{equation}
where $\rho_{in}$ is the initial state and the $A_{i}$ are
matrices known as {\it Kraus} operators (see part (a) of Fig.
\ref{fig1} for illustration).
The normalisation of probabilities implies that Kraus operators
must satisfy $\sum_i {A_{i}}^{\dagger}{A_{i}}=\id$. In some
situations, for example when a system is interacting with an
environment, all or part of the measurement outcomes might not be
accessible. In the most extreme case this corresponds to the
situation where the ancilla particles are being traced out. Then
the map is given by
\begin{equation}
    \label{eq2}
    \sigma = \sum_i A_i \rho_{in} A_i^{\dagger}
\end{equation}
which is illustrated in part (b) of Fig. (\ref{fig2}).
Such a map is often referred to as a {\it trace preserving}
quantum operation, whereas operations in which measurement
outcomes are retained are sometimes referred to as {\it measuring}
quantum operations (or sometimes also {\it selective} quantum
operations, or {\it stochastic} quantum operations, depending upon
the context). Conversely, it can be shown (see e.g. \cite{Nielsen
C 00}) that for {\it any} set of linear operators $A_i$ satisfying
$\sum_i A_i^{\dagger}A_i=\id$ we can find a process, composed of
the addition of ancillae, joint unitary evolution, and von-Neumann
measurements, that leads to eq. (\ref{eq1}). In trace preserving
operations the $A_i$ should strictly all be matrices of the same
dimensions, however, if knowledge of outcomes is retained, then
different $A_i$ may have different dimensions.
Having summarized the basic ingredients of generalised quantum
operations, we are in a position to consider approaches that may
be taken to determine which operations are implementable by LOCC.
The LOCC constraint is illustrated in figure \ref{fig2}. In
general this set of operations is quite complicated. Alice and Bob
may communicate classically before or after any given round of
local actions, and hence in any given round their actions may
depend upon the outcomes of previous measuring operations. As a
consequence of this complexity, there is no known simple
characterisation of the LOCC operations. This has motivated the
development of larger classes of operations that can be more
easily characterised, while still retaining a considerable element
of LOCC-ality. One of the most important such classes is the set
of {\it separable operations}. These are the operations that can
be written in terms of Kraus operators with a {\it product}
decomposition:
\begin{equation}
    \label{eq3}
    \rho_{k} = \frac{ A_k\otimes B_k \rho_{in} A_k^{\dagger}\otimes B_k^{\dagger}}
    {tr A_k \otimes B_k \rho_{in} A_k^{\dagger}\otimes B_k^{\dagger}}
\end{equation}
such that $\sum_k A_k^{\dagger}A_k\otimes B_k^{\dagger} B_k
=\id\otimes\id$. Clearly, any LOCC operation can be cast in the
form of separable operation, as the local Kraus operators
corresponding to the individual actions of Alice and Bob can be
joined into product Kraus operators. However, it is remarkable
that the converse is {\it not} true. This was first demonstrated
in \cite{Bennett DFMRSS 99}, where an example task of a separable
operation is presented that cannot be implemented using LOCC
actions - the example presented there requires a finite amount of
quantum communication to implement it, even though the operation
is itself separable.

It is nevertheless convenient from a mathematical point of view to
work with separable operations, as optimising a given task using
separable operations provides strong bounds on what may be
achieved using LOCC. Sometimes this process can even lead to tight
results - one may try to show whether the optimal separable
operation may in fact be also implemented using LOCC, and this can
often, but not always, be guaranteed in the presence of symmetries
(see e.g. \cite{Rains 01,Virmani P 03} and refs. therein). Even
more general classes of operations such as positive partial
transpose preserving operations (PPT) \footnote{The class of PPT
operations was proposed by Rains \cite{Rains 99,Rains 01}, and is
defined as the set of completely positive operations $\Phi$ such
that $\Gamma_B \circ \Phi \circ \Gamma_B$ is also completely
positive, where $\Gamma_B$ corresponds to transposition of {\it
all} of Bob's particles, including ancillas. One can also consider
transposition only of those particles belonging to Bob that
undergo the operation $\Phi$. However, we believe that this does
not affect the definition. It is also irrelevant whether the
transposition is taken over Alice or Bob, and so one may simply
assert that $\Gamma \circ \Phi \circ \Gamma$ must be completely
positive, where $\Gamma$ is the transposition of one party. It can
be shown that the PPT operations are precisely those operations
that preserve the set of PPT states. Hence the set of non-PPT
operations includes any operation that creates a free (non-bound)
entangled state out of one that is PPT. Hence PPT operations
correspond to some notion of locality, and in contrast to
separable operations it is relatively easy to check whether a
quantum operation is PPT \cite{Rains 01}.} may also be used in the
study of entanglement as they have the advantage of a very compact
mathematical characterization \cite{Rains 01,Eggeling VWW
01,Audenaert PE 03}.

After this initial discussion of quantum operations and the LOCC
constraint we are now in a position to consider in more detail the
basic properties of entanglement.

{\bf\em Basic properties of entanglement --} Following our
discussion of quantum operations and their natural constraint to
local operations and classical communication, we are now in a
position to establish some basic facts and definitions regarding
entangled states. Given the wide range of tasks that exploit
entanglement one might try to define entanglement as `that
property which is exploited in such protocols'. However, there is
a whole range of such tasks, with a whole range of possible
measures of success. This means that situations will almost
certainly arise where a state $\rho_1$ is better than another
state $\rho_2$ for achieving one task, but for achieving a
different task $\rho_2$ is better than $\rho_1$. Consequently
using a task-based approach to quantifying entanglement will
certainly not lead to a single unified perspective. However,
despite this problem, it is possible to assert some general
statements which are valid regardless of what your favourite use
of entanglement is, as long as the key set of `allowed' operations
is the LOCC class. This will serve us a guide as to how to
approach the quantification of entanglement, and so we will
discuss some of these statements in detail:

\smallskip

$\bullet$ {\it Separable states contain no entanglement.}

A state $\rho_{ABC...}$ of many parties $A,B,C,...$ is said to be
{\it separable} \cite{Werner 89}, if it can be written in the form
\begin{equation}
    \label{separable}
    \rho_{ABC...} = \sum_i p_i ~ \rho^i_{A} \otimes \rho^i_{B} \otimes
    \rho^i_{C} \otimes ...
\end{equation}
where $p_i$ is a probability distribution. These states can
trivially be created by LOCC - Alice samples from the distribution
$p_i$, informs all other parties of the outcome $i$, and then each
party $X$ locally creates $\rho^i_X$ and discards the information
about the outcome $i$. As these states can be created from scratch
by LOCC they trivially satisfy a local hidden variables model and
all their correlations can be described classically. Hence, it is
quite reasonable to state that separable states contain no
entanglement.

\smallskip

$\bullet$ {\it All non-separable states allow some tasks to be
achieved better than by LOCC alone, hence all non-separable states
are entangled.}

For a long time the quantum information community has used a
`negative' characterization of the term entanglement essentially
defining entangled states as those that cannot be created by LOCC
alone. On the other hand, it can be shown that a quantum state
$\rho$ may be generated perfectly using LOCC if and only if it is
separable. Of course this is a task that becomes trivially possible
by LOCC when the state $\rho$ has been provided as a non-local
resource in the first place. More interestingly, it has been shown
recently that for any non-separable state $\rho$, one can find
another state $\sigma$ whose teleportation fidelity may be enhanced
if $\rho$ is also present \cite{Masanes 05,Masanes 05a,Brandao 05}.
This is interesting as it allows us to positively characterize
non-separable states as those possessing a useful resource that is
not present in separable states. This hence justifies the synonymous
use of the terms {\it non-separable} and {\it entangled}.
\smallskip

$\bullet$ {\it The entanglement of states does not increase under
LOCC transformations.}

Given that by LOCC we can only create separable, ie non-entangled
states, this immediately implies the statement that LOCC cannot
create entanglement from an unentangled state. Indeed, we even
have the following stronger fact. Suppose that we know that a
quantum state $\rho$ can be transformed with certainty to another
quantum state $\sigma$ using LOCC operations. Then anything that
we can do with $\sigma$ and LOCC operations we can also achieve
with $\rho$ and LOCC operations. Hence the utility of quantum
states cannot increase under LOCC operations \cite{Bennett BPS
96,Vedral PRK 97,Vedral P 98,Plenio V 98}, and one can rightfully
state that $\rho$ is at least as entangled as $\sigma$.

\smallskip

$\bullet$ {\it Entanglement does not change under Local Unitary
operations.}

This property follows from the previous one because local unitaries
can be inverted by local unitaries. Hence, by the non-increase of
entanglement under LOCC, two states related by local unitaries have
an equal amount of entanglement.

\smallskip

$\bullet$ {\it There are maximally entangled states.}

Now we have a notion of which states are entangled and are also
able, in some cases, to assert that one state is more entangled
than another. This naturally raises the question whether there is
a {\it maximally entangled state}, i.e. one that is more entangled
than all others. Indeed, at least in two-party systems consisting
of two fixed $d$-dimensional sub-systems (sometimes called
qudits), such states exist. It turns out that any pure state that
is local unitarily equivalent to
\begin{displaymath}
    |\psi_d^+\rangle= \frac{|0,0\rangle + |1,1\rangle + .. + |d-1,d-1\rangle}{\sqrt{d}}
\end{displaymath}
is maximally entangled. This is well justified, because as we shall
see in the next subsection, any pure or mixed state of two
$d$-dimensional systems can be prepared from such states with
certainty using only LOCC operations. We shall later also see that
the non-existence of an equivalent statement in multi-particle
systems is one of the reasons for the difficulty in establishing a
theory of multi-particle entanglement.
\\

\smallskip

The above considerations have given us the extremes of
entanglement - as long as we consider LOCC as our set of available
operations, separable states contain zero entanglement, and we can
identify certain states that have maximal entanglement. They also
suggest that we can impose some form of ordering - we may say that
state $\rho$ is more entangled than a state $\sigma$ if we can
perform the transformation $\rho \rightarrow \sigma$ using LOCC
operations. A key question is whether this method of ordering
gives a partial or total order? To answer this question we must
try and find out when one quantum state may be transformed to
another using LOCC operations. Before we move on to the discussion
of entanglement measures we will consider this question in more
detail in the next part.

Note that the notion that `{\it entanglement does not increase under
LOCC}' is implicitly related to our restriction of quantum
operations to LOCC operations - if other restrictions apply, weaker
or stronger, then our notion of `more entangled' is likely to also
change.

\section{Local Manipulation of Quantum States}

{\em\bf Manipulation of single bi-partite states --} In the
previous section we indicated that for bi-partite systems there is
a notion of maximally entangled states that is independent of the
specific quantification of entanglement. This is so because there
are so-called {\it maximally entangled states} from which all
others can be created by LOCC only (at least for bipartite systems
of fixed maximal dimension). We we will show this explicitly here
for the case of two qubits and leave the generalization as an
exercise to the reader. In the case of two qubits, we will see
that the maximally entangled states are those that are
local-unitarily equivalent to the state
\begin{equation}
    |\psi_{2}^+\rangle = \frac{1}{\sqrt{2}}(|00\rangle +
    |11\rangle)\, .
\end{equation}
Our aim is now to justify this statement by showing that for any
bipartite pure state written in a Schmidt decomposed form (see
discussion around equation (\ref{schmidt}) for an explanation of the
Schmidt Decomposition):
\begin{equation}
    |\phi\rangle = \alpha|00\rangle + \beta|11\rangle
\end{equation}
we can find a LOCC map that takes $|\psi_2^+\rangle$ to
$|\phi\rangle$ with certainty. To this end we simply need to write
down the Kraus operators (see eq. (\ref{eq1}) of a valid quantum
operation. It is easy to show that the Kraus operators defined by
\begin{eqnarray}
    A_0 &:=& (\alpha|0\rangle\langle 0| + \beta |1\rangle\langle
    1|)\otimes \id, \nonumber\\
    A_1 &:=& (\beta|1\rangle\langle 0| + \alpha |0\rangle\langle 1|)
    \otimes (|1\rangle\langle 0| + |0\rangle\langle 1|)
\end{eqnarray}
 satisfy $A_0^{\dagger}A_0+A_1^{\dagger}A_1 = \id\otimes\id$ and
$A_i|\psi\rangle = p_i|\phi\rangle$, so that
$|\phi\rangle\langle\phi|
=
A_0|\psi\rangle\langle\psi|A_0^{\dagger}+A_1|\psi\rangle\langle\psi|A_1^{\dagger}$.
It is instructive to see how one can construct this operation
physically using only LOCC transformations. Let us first add an
ancilla in state $|0\rangle$ to Alice which results in the state
\begin{equation}
    \frac{|00\rangle_A|0\rangle_B +
    |01\rangle_A|1\rangle_B}{\sqrt{2}}\, .
\end{equation}
If we then perform the local unitary operation $|00\rangle
\rightarrow \alpha|00\rangle + \beta|11\rangle;\, |01\rangle
\rightarrow \beta|01\rangle + \alpha|10\rangle $ on Alice's two
particles, we arrive at
\begin{equation}
    \frac{|0\rangle_A(\alpha|00\rangle_{AB}+\beta|11\rangle_{AB})
    +
    |1\rangle_A(\beta|10\rangle_{AB}+\alpha|01\rangle_{AB})}{\sqrt{2}}.
\end{equation}
Finally, a local measurement on Alice's ancilla particle now
yields two outcomes. If Alice finds $|0\rangle$ then Bob is
informed and does not need to carry out any further operation; if
Alice finds $|1\rangle$ then Bob needs to apply a $\sigma_x$
operation to his particle. In both cases this results in the
desired state $\alpha|00\rangle_{AB}+\beta|11\rangle_{AB}$.

Given that we can obtain with certainty any arbitrary pure state
starting from $|\psi_2^+\rangle$, we can also obtain any mixed state
$\rho$. This is because any mixed state $\rho$ can always be written
in terms of its eigenvectors as $\rho = \sum_{i} p_i
|\phi_i\rangle\langle\phi_i|$, where each eigenvector is of the form
$|\phi_i\rangle = U_i\otimes V_i (
\alpha_i|00\rangle+\beta_i|11\rangle)$ for some set of unitaries
$U_i$ and $V_i$ (this in turn is simply a consequence of the Schmidt
decomposition). It is an easy exercise, left to the reader, to
construct the operation that takes $|\psi_2^+\rangle$ to $\rho$.

A natural generalisation of this observation would be to consider
LOCC transformations between general pure states of two parties
\cite{Lo P 01}. Although this question is a little more difficult,
a complete solution has been developed using the mathematical
framework of the theory of {\it majorization}. The results that
have been obtained not only provide necessary and sufficient
conditions for the possibility of the LOCC interconversion between
two pure states, they are also constructive as they lead to
explicit protocols that achieve the task \cite{Nielsen 99,Vidal
99,Jonathan P 99a,Hardy 99}. These conditions may be expressed
most naturally in terms of the {\it Schmidt coefficients}
\cite{Nielsen C 00} of the states involved. It is a useful fact
that any bi-partite pure quantum state $|\psi\rangle$ may be
written in the form
\begin{equation}
    |\psi\rangle = U_A\otimes U_B \sum_{i=1}^{N}
    \sqrt{\alpha_i}|i\rangle_A|i\rangle_B \label{schmidt}
\end{equation}
where the positive real numbers $\alpha_i$ are the {\it
Schmidt-coefficients} of the state $|\psi\rangle$ \footnote{That
this is true can be proven as follows. Consider a general
bipartite state $|\psi\rangle = \sum a_{ij}|i\rangle|j\rangle$.
The amplitudes $a_{ij}$ can be considered as the matrix elements
of a matrix $A$. This matrix hence completely represents the state
(as long as a local basis is specified). If we perform the local
unitary transformation $U \otimes V|\psi\rangle$ then the matrix
$A$ gets transformed as $A \rightarrow UAV^T$. It is a well
established result of matrix analysis - the {\it singular value
decomposition} \cite{Bhatia 97} - that any matrix $A$ can be {\it
diagonalised} into the form $A_{ij}=\lambda_{i}\delta_{ij}$ by a
suitable choice of ($U,V$), even if $A$ is not square. The
coefficients $\lambda_i$ are the so-called {\it singular values}
of $A$, and correspond to the Schmidt coefficients.}. The local
unitaries do not affect the entanglement properties, which is why
we now write the initial state vector $|\psi_1\rangle$ and final
state vector $|\psi_2\rangle$ in their Schmidt-bases,
\begin{eqnarray}
    \left| \psi_1 \right\rangle = \sum^{n}_{i=1}
    \sqrt{\alpha_i}\left|i_A \right\rangle \left|i_B
    \right\rangle,~~~
    \left| \psi_2 \right\rangle = \sum^{n}_{i=1}\sqrt{\alpha'_i}
    \left|i_A' \right\rangle \left|i_B' \right\rangle \nonumber
\end{eqnarray}
where $n$ denotes the dimension of each of the quantum systems. We
can take the Schmidt coefficients to be given in decreasing order,
i.e., $\alpha_1\ge\alpha_2\ge \ldots\ge \alpha_n$ and
$\alpha'_1\ge\alpha'_2\ge \ldots\ge \alpha'_n$. The question of the
interconvertibility between the states can then be decided from the
knowledge of the real Schmidt coefficients only, as any two pure
states with the same Schmidt coefficients may be interconverted
straightforwardly by local unitary operations. In \cite{Nielsen 99}
it has been shown that a LOCC transformation converting
$\left|\psi_1\right\rangle $ to $\left| \psi_2 \right\rangle $ with
unit probability exists if and only if the $\left\{\alpha_i
\right\}$ are  {\it majorized} by $\left\{\alpha'_i\right\}$, i.e.
if for all $1\leq l< n$ we have that
\begin{equation}\label{majorization}
    \sum_{i=1}^{l}\alpha _{i}\leq \sum_{i=1}^{l}\alpha _{i}^{\prime }
\end{equation}
and $\sum_{i=1}^n \alpha_i = \sum_{i=1}^n \alpha_i'$, where $n$
denotes the number of nonzero Schmidt-coefficients \cite{Bhatia
97}. Various refinements of this result have been found that
provide the largest success probabilities for the interconversion
between two states by LOCC, together with the optimal protocol
(according to certain figures of merit) where such a deterministic
interconversion is not possible \cite{Vidal 99,Jonathan P
99a,Jonathan P 99b}. These results allow us in principle to decide
any question concerning the LOCC interconversion of pure states by
employing techniques from linear programming \cite{Jonathan P
99a}.

It is a direct consequence of the above structures that there are
{\em incomparable} states, i.e. pairs of states such that neither
can be converted into the other with certainty. These states are
called incomparable as neither can be viewed as more entangled than
the other. Note that borrowed entanglement can make some pairs of
incomparable states comparable again. Indeed, there are known
examples where the LOCC transformation of
$|\psi\rangle\rightarrow|\phi\rangle$ is not possible with
probability one, but where given a suitable entangled state
$|\eta\rangle$ the LOCC transformation of
$|\psi\rangle|\eta\rangle\rightarrow|\phi\rangle|\eta\rangle$ is
possible with certainty \cite{Jonathan P 99b}. This phenomenon is
now called {\it entanglement catalysis}, as the state $|\eta\rangle$
is returned unchanged after the transformation, and acts much like a
catalyst. The majorization condition also reveals another
disadvantageous feature of the single copy setting - there can be
{\it discontinuities}. For instance, it can be shown that the
maximal probability of success for the LOCC transformation from
$(|00\rangle+|11\rangle)/\sqrt{2}$ to $0.8|00\rangle+0.6|11\rangle$
is unity, while the probability for the transformation
$(|00\rangle+|11\rangle)/\sqrt{2}$ to
$(0.8|00\rangle+0.6|11\rangle+\epsilon
|22\rangle)/\sqrt{1+\epsilon^2}$ is strictly zero for any
$\epsilon\neq 0$, i.e. even if the target states in the two examples
are arbitrarily close. That the probability of success for the later
transformation is zero can also be concluded easily from the fact
that the Schmidt-number, i.e. the number of non-vanishing
Schmidt-coefficients, cannot be increased in an LOCC protocol, even
probabilistically.

The key problem is that we are being too restrictive in asking for
{\em exact} state transformations. Physically, we should be
perfectly happy if we can come very close to a desired state.
Indeed, admitting a small but finite value of $\epsilon$ there
will be a finite probability to achieve the transformation. This
removes the above discontinuity \cite{Vidal JN 00}, but the
success probability will now depend on the size of the imprecision
that we allow. The following subsection will serve to overcome
this problem for pure states by presenting a natural definition of
interconvertibility in the presence of vanishing imprecisions, a
definition that will constitute our first entanglement measure.

{\em\bf State manipulation in the asymptotic limit --} The study
of the LOCC transformation of pure states has so far enabled us to
justify the concept of maximally entangled states and has also
permitted us, in some cases, to assert that one state is more
entangled than another. However, we know that exact LOCC
transformations can only induce a partial order on the set of
quantum states. The situation is even more complex for {\it mixed}
states, where even the question of when it is possible to LOCC
transform one state into another is a difficult problem with no
transparent solution at the time of writing.

All this means that if we want to give a definite answer as to
whether one state is more entangled than another for any pair of
states, it will be necessary to consider a more general setting.
In this context a very natural way to compare and quantify
entanglement is to study LOCC transformations of states in the so
called {\it asymptotic regime}. Instead of asking whether for a
single pair of particles the initial state $\rho$ may be
transformed to a final state $\sigma$ by LOCC operations, we may
ask whether for some large integers $m,n$ we can implement the
`wholesale' transformation $\rho^{\otimes n} \rightarrow
\sigma^{\otimes m}$.
The largest ratio $m/n$ for which one may achieve this would then
indicate the relative entanglement content of these two states.
Considering the many-copy setting allows each party to perform
collective operations on (their shares of) many copies of the states
in question. Such a many copy regime provides many more degrees of
freedom, and in fact paves part of the way to a full classification
of pure entangled states. To pave the rest of the route we will also
need to discuss what kind of approximations we might admit for the
output of the transformations.

There are two basic approaches to this problem - we can consider
either {\it exact} or {\it asymptotically exact} transformations.
The distinction between these two approaches is important, as they
lead to different scenarios that yield different answers. In the
{\it exact} regime we allow no errors at all - we must determine
whether the transformation $\rho^{\otimes n} \rightarrow
\sigma^{\otimes m}$ can be achieved perfectly and with $100\%$
success probability for a given value of $m$ and $n$. The supremum
of all achievable rates $r=m/n$ is denoted by
$r_{exact}(\sigma\rightarrow\rho)$, and carries significance as a
measure of the exact LOCC `exchange rate' between states
$\rho,\sigma$. This quantity may be explored and gives some
interesting results \cite{Audenaert PE 03}. However, from a
physical point of view one may feel that the restriction to exact
transformations is too stringent. After all, it should be quite
acceptable to consider approximate transformations \cite{Bennett
BPS 96} that become arbitrarily precise when going to the
asymptotic limit. Asymptotically vanishing imperfections, as
quantified by the trace norm (i.e. {tr$|\sigma - \eta|$}), will
lead to vanishingly small changes in measurements of bounded
observables on the output. This leads to the second approach to
approximate state transformations, namely that of {\it
asymptotically exact} state transformations, and this is the
setting that we will consider for the remainder of this work.
In this setting we consider imperfect transformations between
large blocks of states, such that in the limit of large block
sizes the imperfections vanish. For example, for a large number
$n$ of copies of $\rho$, one transforms $\rho^{\otimes n}$ to an
output state $\sigma_m$ that approximates $\sigma^{\otimes m}$
very well for some large $m$. If, in the limit of
$n\rightarrow\infty$ and for fixed $r=m/n$, the approximation of
$\sigma^{\otimes m}$ by $\sigma_m$ becomes arbitrarily good, then
the rate $r$ is said to be {\it achievable}. One can use the
optimal (supremal) achievable rate $r_{approx}$ as a measure of
the relative entanglement content of $\rho,\sigma$ in the
asymptotic setting. This situation is reminiscent of Shannon
compression in classical information theory - where the
compression process loses all imperfections in the limit of
infinite block sizes as long as the compression rate is below a
threshold \cite{Cover T}. Clearly the asymptotically exact regime
is less strongly constrained than the exact regime, so that
$r_{approx}\ge r_{exact}$. Given that we are considering two
limiting processes it is not clear whether the two quantities are
actually equal and it can be rigorously demonstrated that they are
different in general, see e.g. \cite{Audenaert PE 03}.

Such an asymptotic approach will alleviate some of the problems
that we encountered in the previous section. It turns out that the
asymptotic setting yields a unique total order on bi-partite pure
states, and as a consequence, leads to a very natural measure of
entanglement that is essentially unique. To this end let us start
by defining our first entanglement measure, which happens also to
be one of the most important measures - the {\it entanglement
cost}, $E_C (\rho)$. For a given state $\rho$ this measure
quantifies the maximal possible rate $r$ at which one can convert
blocks of {\it 2-qubit} maximally entangled states
\cite{footnote2} into output states that approximate many copies
of $\rho$, such that the approximations become vanishingly small
in the limit of large block sizes. If we denote a general trace
preserving LOCC operation by $\Psi$, and write $\Phi(K)$ for the
density operator corresponding to the maximally entangled state
vector in $K$ dimensions, i.e.\
$\Phi(K)=|\psi_K^+\rangle\langle\psi_K^+|$, then the entanglement
cost may be defined as
\begin{displaymath}
    E_C(\rho) = \inf\left\{ r: \lim_{n\rightarrow\infty}
    \left[\inf_{\Psi} D(\rho^{\otimes n},\Psi(\Phi(2^{rn})))\right] = 0 \right\}
\end{displaymath}
where D$(\sigma,\eta)$ is a suitable measure of distance
\cite{Rains 97,Audenaert PE 03}. A variety of possible distance
measures may be considered. It has been shown that the definition
of entanglement cost is independent of the choice of distance
function, as long as these functions are equivalent to the trace
norm in a way that is sufficiently independent of dimension (see
\cite{Hayden HT 01} for further explanation). Hence we will fix
the trace norm distance, i.e. $D(\sigma,\eta)=tr|\sigma-\eta|$, as
our canonical choice of distance function.

It may trouble the reader that in the definition of $E_C(\rho)$ we
have not actually taken input states that are blocks of $rn$
copies of 2-qubit maximally entangled states, but instead have
chosen as inputs single maximally entangled states between
subsystems of increasing dimensions $2^{rn}$. However, these two
approaches are equivalent because (for integral $rn$)
$\Phi(2^{rn})$ is local unitarily equivalent to $\Phi(2)^{\otimes
rn}$.

The entanglement cost is an important measure because it
quantifies a wholesale `exchange rate' for converting maximally
entangled states to $\rho$ by LOCC alone. Maximally entangled
states are in essence the `gold standard currency' with which one
would like to compare all quantum states. Although computing
$E_C(\rho)$ is extremely difficult, we will later discuss its
important implications for the study of channel capacities, in
particular via another important and closely related entanglement
measure known as the {\it entanglement of formation}, $E_F(\rho)$.

Just as $E_C(\rho)$ measures how many maximally entangled states
are required to create copies of $\rho$ by LOCC alone, we can ask
about the reverse process: at what rate may we obtain maximally
entangled states (of two qubits) from an input supply of states of
the form $\rho$. This process is known in the literature either as
{\it entanglement distillation}, or as {\it entanglement
concentration} (usually reserved for the pure state case). The
efficiency with which we can achieve this process defines another
important basic asymptotic entanglement measure which is the {\it
Distillable Entanglement}, $E_D(\rho)$. Again we allow the output
of the procedure to {\em approximate} many copies of a maximally
entangled state, as the exact transformation from $\rho^{\otimes
n}$ to even one pure maximally entangled state is in general
impossible \cite{Kent}. In analogy to the definition of
$E_C(\rho)$, we can then make the precise mathematical definition
of $E_D(\rho)$ as
\begin{displaymath}
    E_D(\rho) := \sup\left\{ r: \lim_{n\rightarrow\infty}
    \left[\inf_{\Psi} {\rm tr}|\Psi(\rho^{\otimes n})-\Phi(2^{rn})|\right] =
    0 \right\}.
\end{displaymath}
$E_D(\rho)$ is an important measure because if entanglement is used
in a two party quantum information protocol, then it is usually
required in the form of maximally entangled states. So $E_D(\rho)$
tells us the rate at which noisy mixed states may be converted back
into the `gold standard' singlet state by LOCC. In defining
$E_D(\rho)$ we have ignored a couple of important issues. Firstly,
our LOCC protocols are always taken to be trace preserving. However,
one could conceivably allow probabilistic protocols that have
varying degrees of success depending upon various measurement
outcomes. Fortunately, a thorough analysis by Rains \cite{Rains 99}
shows that taking into account a wide diversity of possible success
measures still leads to the same notion of distillable entanglement.
Secondly, we have always used 2-qubit maximally entangled states as
our `gold standard'. If we use other entangled {\it pure} states,
perhaps even on higher dimensional Hilbert spaces, do we arrive at
significantly altered definitions? We will very shortly see that
this is not the case so there is no loss of generality in taking
singlet states as our target.

Given these two entanglement measures it is natural to ask whether
$E_C{?\atop =}E_D$, i.e. whether entanglement transformations become
{\em reversible} in the asymptotic limit. This is indeed the case
for pure state transformations where $E_D(\rho)$ and $E_C(\rho)$ are
identical and equal to the {\em entropy of entanglement}
\cite{Bennett BPS 96}. The entropy of entanglement for a pure state
$|\psi\rangle$ is defined as
\begin{equation} \label{entropyofentanglement}
    E(|\psi\rangle\langle \psi|): = S(\mbox{tr}_A
    |\psi\rangle\langle\psi|) = S(\mbox{tr}_B
    |\psi\rangle\langle\psi|)
\end{equation}
where $S$ denotes the von-Neumann entropy $S(\rho)=-\mbox{tr}[
\rho\log_2 \rho]$, and $\mbox{tr}_B$ denotes the partial trace
over subsystem B. This reversibility means that in the asymptotic
regime we may immediately write down the optimal rate of
transformation between {\it any} two pure states $|\psi_1\rangle$
and $|\psi_2\rangle$. Given a large number $N$ of copies of
$|\psi_1\rangle\langle \psi_1|$, we can first distill $\approx
NE(|\psi_1\rangle\langle \psi_1|)$ singlet states and then create
from those singlets $M \approx N E(|\psi_1\rangle\langle
\psi_1|)/E(|\psi_2\rangle\langle \psi_2|)$ copies of
$|\psi_2\rangle\langle \psi_{2}|$. In the infinite limit these
approximations become exact, and as a consequence
$E(|\psi_1\rangle\langle \psi_1|)/E(|\psi_2\rangle\langle
\psi_2|)$ is the optimal asymptotic conversion rate from
$|\psi_1\rangle\langle \psi_{1}|$ to $|\psi_2\rangle\langle
\psi_{2}|$. It is the reversibility of pure state transformations
that enables us to define $E_D(\rho)$ and $E_C(\rho)$ in terms of
transformations to or from singlet states - the use of any other
entangled pure state (in any other dimensions) simply leads to a
constant factor multiplied in front of these quantities.

Following these basic considerations we are now in a position to
formulate a more rigorous and axiomatic approach to entanglement
measures that captures the lessons that have been learned in the
previous sections. In the final section we will then review
several entanglement measures, presenting useful formulae and
results and discuss the significance of these measures for various
topics in quantum information.\\

\section{Postulates for axiomatic Entanglement Measures}

In the previous section we considered the quantification of
entanglement from the perspective of LOCC transformations in the
asymptotic limit. This approach is interesting because it can be
solved completely for pure states. It demonstrates that LOCC
entanglement manipulation is reversible in this setting, therefore
imposing a unique order on pure entangled states via the entropy of
entanglement. However, for mixed states and LOCC operations the
situation is more complicated and reversibility is lost \cite{Vidal
C 01,Horodecki SS 02}.

The concomitant loss of a total
ordering of quantum states (in terms of rates of LOCC entanglement
interconversions) implies that in general an LOCC based
classification of entanglement would be extremely complicated.

However, one can try to salvage the situation by taking a more
axiomatic approach. One can {\it define} real valued functions that
satisfy the basic properties of entanglement that we outlined
earlier, and use these functions to attempt to {\it quantify} the
amount of entanglement in a given quantum state. This is essentially
the process that is followed in the definition of most entanglement
measures. Various such quantities have been proposed over the years,
such as the entanglement of distillation \cite{Bennett BPS 96,Rains
99}, the entanglement cost \cite{Bennett BPS 96,Bennett DSW
96,Wootters 98,Hayden HT 01,Wootters 01}, the relative entropy of
entanglement \cite{Vedral PRK 97,Vedral PJK 97,Vedral P 98} and the
squashed entanglement \cite{Christandl W 03}. Some of these measures
have direct physical significance, whereas others are purely
axiomatic. Initially these measures were used to give a physically
motivated classification of entanglement that is simple to
understand, and can even be used to assess the quality of entangled
states produced in experiments. However, subsequently they have also
been developed into powerful mathematical tools, with great
significance for open questions such as the additivity of quantum
channel capacities \cite{Shor,Audenaert B 04,Pom}, quantifying
quantum correlations in quantum-many-body systems \cite{Verstraete
DC 04,Pachos P 04,Verstraete PC 04,Popp VMC 05,Key LPPRR 05}, and
bounding quantum computing fault tolerance thresholds \cite{Harrow
N,Virmani HP 05} to name a few.

In this section we will now discuss and present a few basic axioms
that any measure of entanglement should satisfy. This will allow us
to define further quantities that go beyond the two important mixed
state measures ($E_C(\rho)$ and $D(\rho)$) that we have already
introduced.

So what exactly are the properties that a good entanglement measure
should possess? An entanglement measure is a mathematical quantity
that should capture the essential features that we associate with
entanglement, and ideally should be related to some operational
procedure. Depending upon your aims, this can lead to a variety of
possible desirable properties. The following is a list of possible
postulates for entanglement measures, some of which are not
satisfied by all proposed quantities \cite{Vedral P 98,Donald HR
01}:

\begin{enumerate}
    \item A {\it bipartite} entanglement measure $E(\rho)$ is a mapping from density
    matrices into positive real numbers:
    \begin{equation}
    \rho \rightarrow E(\rho) \in \mathbb{R}^+
    \end{equation}
    defined for states of arbitrary bipartite systems. A
    normalisation factor is also usually included
    such that the maximally entangled state
    $$|\psi_d^+\rangle= \frac{|0,0\rangle + |1,1\rangle + .. +|d-1,d-1\rangle}{\sqrt{d}}$$
    of two qudits has
    $E(|\psi_d^+\rangle)=\log d$.
    \item $E(\rho)=0$ if the state $\rho$ is separable.
    \item $E$ does not increase on average
    under LOCC, i.e.,
    \begin{equation}
        E(\rho) \ge \sum_{i} p_i E(\frac{A_i\rho A_i^{\dagger}}{tr A_i\rho A_i^{\dagger}})
    \end{equation}
    where the $A_i$ are the Kraus operators describing some LOCC protocol
    and the probability of obtaining outcome $i$ is given by $p_i=tr A_i\rho A_i^{\dagger}$
    (see fig \ref{fig1}).
    \item  For pure state $|\psi\rangle\langle\psi|$ the measure
    reduces to the entropy of entanglement
    \begin{equation}
        E(|\psi\rangle\langle\psi|) = (S\circ \mbox{tr}_B)
    (|\psi\rangle\langle\psi|) ~ .
    \end{equation}
\end{enumerate}
 We will call any function $E$ satisfying the first
three conditions an {\it entanglement monotone}. The term {\it
entanglement measure} will be used for any quantity that satisfies
axioms 1,2 and 4, and also does not increase under {\it
deterministic} LOCC transformations. In the literature these terms
are often used interchangeably.
Note that the conditions (1)-(4) may be replaced by an
equivalent set of slightly more abstract conditions which will be
explained below eq. (\ref{continuity20}). Frequently, some authors
also impose additional requirements for entanglement measures:

$\bullet$ {\em Convexity --} One common example for an additional
property required from an entanglement measure is the concept of
convexity which means that we require
\begin{displaymath}
    E(\sum_i p_i \rho_i) \le \sum_i p_i E(\rho_i).
\end{displaymath}
Requiring this mathematically very convenient property is sometimes
justified as capturing the notion of the loss of information, i.e.
describing the process of going from a selection of identifiable
states $\rho_i$ that appear with rates $p_i$ to a mixture of these
states of the form $\rho=\sum p_i \rho_i$. We would like to stress,
however, that some care has to be taken in this context. Indeed, in
the first situation, when the states are locally identifiable, the
whole ensemble can be described by the quantum state
\begin{equation}
    \sum_i p_i |i\rangle_M\langle i|\otimes \rho_i^{AB},
\end{equation}
where the $\{|i\rangle_M\}$ denote some orthonormal basis for a
particle belonging to one of the two parties. Clearly a measurement
of the marker particle $M$ reveals the identity of the state of
parties $A$ and $B$. Losing the association between $|i\rangle_M$
and state $\rho_i^{AB}$ then correctly describes the process of the
forgetting, a process which is then described by tracing out the
marker particle $M$ to obtain $\rho=\sum p_i \rho_i$ \cite{Plenio V
01,Plenio 05}. As this is a local operation we may then require that
$E(\sum_i p_i |i\rangle_M\langle i|\otimes \rho_i^{AB}) \ge
E(\rho)$, which is, of course, already required by condition 3
above. Hence there is no strict need to require convexity, except
for the mathematical simplicity that it might bring. A compelling
example of the technical simplicity that convexity can bring is the
very simple test for entanglement monotonicity of a convex function
$f$. Indeed, a convex function $f$ does not increase under LOCC if
and only if it satisfies (i) $f(U_A\otimes U_B \rho_{AB}
U_A^{\dagger}\otimes U_B^{\dagger})=f(\rho_{AB})$ for all local
unitary $U_A,U_B$ and (ii) $f(\sum_i p_i \rho_{AB}^{i} \otimes
|i\rangle\langle i|_X)=\sum_i p_i f(\rho_{AB}^{i})$ where $X=A',B'$
and $|i\rangle$ form local, orthogonal bases \cite{Horodecki 04}.

$\bullet$ {\em Additivity --} Given an entanglement measure and a
state $\sigma$ one may ask for the condition $E(\sigma^{\otimes
n}) = nE(\sigma)$ to be satisfied for all integer $n$. A measure
satisfying this property is said to be {\it additive}.
Unfortunately, there are some significant entanglement measures
that do not satisfy this condition, and for this reason we have
not included additivity as a basic postulate. However, given any
measure $E$ that is not additive there is a straightforward way of
removing this deficiency. We may define the {\it regularized}, or
{\it asymptotic} version:
\begin{equation}
    E^{\infty}(\sigma) := \lim_{n\rightarrow\infty} \frac{E(\sigma^{\otimes n})}{n}
\end{equation}
which is a measure that then automatically satisfies additivity.

A much stronger requirement could be to demand {\it full
additivity}, by which we mean that for any pair of states $\sigma$
and $\rho$ we have $E(\sigma\otimes\rho) = E(\sigma) + E(\rho)$.
This is a very strong requirement and in fact it may be too strong
to be satisfied by all quantities that otherwise satisfy the four
basic properties stated above. Indeed, even such basic measures as
the distillable entanglement may not satisfy this property
\cite{Shor ST 00}.
For these reason we have not included the full additivity in our set
of properties. However, additivity can be a very useful mathematical
property, and we will discuss it further in the context of specific
measures.

$\bullet$ {\em Continuity --} Conditions (1-3) listed above seem
quite natural - the first two conditions are little more than
setting the scale, and the third condition is a generalisation of
the idea that entanglement can only decrease under LOCC operations
to incorporate probabilistic transformations. The fourth condition
appears considerably stronger and perhaps arbitrary at first sight.
However, it turns out that it is also quite a natural condition to
impose. In fact we know that $S(\rho_A)$ represents the reversible
rate of conversion between pure states in the asymptotic regime
which strongly suggests that it is the appropriate measure of
entanglement for pure states. This is reinforced by the following
nontrivial observation: it turns out that {\it any} entanglement
monotone that is (a) additive on pure states, and (b) ``sufficiently
continuous'' must {\it equal} $S(\rho_A)$ on all pure states. Before
we see what sufficiently continuous means we present a very rough
argument for this statement. We know from the asymptotic pure state
distillation protocol that from $n$ copies of a pure state
$|\phi\rangle$ we can obtain a state $\rho_n$ that closely
approximates the state $|\psi^-\rangle^{\otimes nE(|\phi\rangle)}$
to within $\epsilon$, where $E(|\phi\rangle)$ is the entropy of
entanglement of $|\phi\rangle$. Suppose therefore that we have an
entanglement monotone $L$ that is {\it additive} on pure states.
Then we may write
\begin{equation}
nL(|\phi\rangle)=L(|\phi\rangle^{\otimes n}) \geq L(\rho_n)
\end{equation}
where the inequality is due to condition 3 for entanglement
monotones. If the monotone $L$ is ``sufficiently continuous'', then
$L(\rho_n)=L(|\psi^-\rangle^{\otimes
nE(|\phi\rangle)})+\delta(\epsilon) =nE(|\phi\rangle) +
\delta(\epsilon)$, where $\delta(\epsilon)$ will be small. Then we
obtain:
\begin{equation}
    L(|\phi\rangle) \geq E(|\phi\rangle)+{\delta(\epsilon) \over
    n}.
\end{equation}
If the function $L$ is sufficiently continuous as the dimension
increases, i.e. ${\delta(\epsilon)/n} \rightarrow 0$ when
$n\rightarrow\infty$, then we obtain $L(|\phi\rangle) \geq
E(|\phi\rangle)$. Invoking the fact that the entanglement cost for
pure states is also given by the entropy of entanglement gives the
reverse inequality $L(|\phi\rangle) \leq E(|\phi\rangle)$ using
similar arguments. Hence sufficiently continuous monotones that are
additive on pure states will naturally satisfy
$L(|\phi\rangle)=E(|\phi\rangle)$. Of course these arguments are not
rigorous, as we have not undertaken a detailed analysis of how
$\delta$ or $\epsilon$ grow with $n$. A rigorous analysis is
presented in \cite{Donald HR 01}, where it is also shown that our
assumptions may be slightly relaxed. The result of this rigorous
analysis is that a function is equivalent to the entropy of
entanglement on pure states {\it if and only if} it is (a)
normalised on the singlet state, (b) additive on pure states, (c)
non-increasing on {\it deterministic} pure state to pure state LOCC
transformations, and (d) {\it asymptotically continuous} on pure
states. The term {\it asymptotically continuous} is defined as the
property
\begin{equation}
    {L(|\phi\rangle_n)-L(|\psi\rangle_n)\over 1 + \log({\rm dim} H_n)}
    \rightarrow 0
    \label{continuity20}
\end{equation}
whenever the trace norm $tr| |\phi\rangle\langle\phi|_n -
|\psi\rangle\langle\psi|_n|$ between two sequences of states
$|\phi\rangle_n,|\psi\rangle_n$ on a sequence of Hilbert spaces
$H_n \otimes H_n$ tends to 0 as $n \rightarrow 0$. It is
interesting to notice that these constraints only concern pure
state properties of $L$, and that they are {\it necessary and
sufficient}. As a consequence of the above discussion we can
conclude that we could have redefined the set of axiomatic
requirements (1)-(4), without changing the set of admissible
measures. We could have replaced axiom (4) with two separate
requirements of (4'a) additivity on pure states and (4'b)
asymptotic continuity on pure states. Together with axiom (3) this
would automatically force any entanglement measure to coincide
with the entropy of entanglement on pure states.

It is furthermore interesting to note that the failure of an
entanglement measure to satisfy asymptotic continuity is strongly
related to the counterintuitive effect of {\em lockability}
\cite{DiVincenzo HLST 03,Horodecki HHO 04,SynakRadtke H 05}. The
basic question behind lockability is: how much can entanglement of
any bi- or multipartite system change when one qubit is discarded? A
measure of entanglement is said to be {\it lockable} if the removal
of a single qubit from a system can reduce the entanglement by an
arbitrarily large amount. This qubit hence acts as a `key' which
once removed `locks' the remaining entanglement. So which
entanglement measures are lockable? The remarkable answer is
this effect can occur for several entanglement measures, including
the Entanglement Cost. On the other hand another class of measures
that will be described later, {\it Relative Entropies of
Entanglement}, are not lockable \cite{Horodecki HHO 04}. It can also
be proven that any measure that is convex and is {\em not}
asymptotically continuous is lockable \cite{Horodecki HHO 04}.

\medskip

{\em Extremal Entanglement Measures--} In the discussions so far we
have formulated several requirements on entanglement measures and
suggested that various measures exist that satisfy those criteria.
It is now interesting to bound the range of such entanglement
measures. One may in fact show that the entanglement cost
$E_C(\rho)$ and the distillable entanglement $E_D(\rho)$ are in some
sense {\it extremal} measures \cite{Horodecki HH 00,Donald HR 01},
in that they are upper and lower bounds for many `{\it wholesale}'
entanglement measures. To be precise, suppose that we have a
quantity $L(\rho)$ satisfying conditions (1) - (3) above, that is
also asymptotically continuous on mixed states, and also has a {\it
regularisation}
\begin{equation}
    \lim_{n \rightarrow \infty} {L(\rho^{\otimes n}) \over n}.
\end{equation}
Then it can be shown that
\begin{equation}
    E_C(\rho) \geq \lim_{n \rightarrow \infty} {L(\rho^{\otimes n})
    \over n} \geq E_D(\rho). \label{upperlower}
\end{equation}
In fact the conditions under which this statement is true are
slightly more general than the ones that we have listed - for more
details see \cite{Donald HR 01}.

\medskip

{\em Entanglement Ordering} The above considerations have allowed
us to impose quite a great deal of structure on entangled states
and the next section will make this even more clear. It should be
noted however that the axioms 1-4 as formulated above are {\em
not} sufficient to give a {\it unique} total ordering in terms of
the entanglement of the set of states. One can show that any two
entanglement measures satisfying axiom 4 can only impose the same
ordering on the set of entangled states if they are actually
exactly the same. More precisely, if for two measures $E_1$ and
$E_2$ and {\em any} pair of states $\sigma_1$ and $\sigma_2$ we
have that $E_1(\sigma_1)\ge E_1(\sigma_2)$ implies
$E_2(\sigma_1)\ge E_2(\sigma_2)$, then if both measures satisfy
axiom 4 it must be the case that $E_1\equiv E_2$ \cite{Virmani P
00} (see \cite{Eisert P 99,Zyczkowski B 02,Miranowicz G 04} for
ordering results for other entanglement quantities). Given that
the entanglement cost and the distillable entanglement are
strictly different on all entangled mixed states \cite{Yang HHS
05} this implies that there is not a unique order, in terms of
entanglement, on the set of entangled states.

This suggests one of several viewpoints. We may for example have
neglected to take account of the resources in entanglement
manipulation with sufficient care, and doing so might lead to the
notion of a unique total order and therefore a unique entanglement
measure \cite{Horodecki HHOSSS 04}. Alternatively, it may be
possible that the setting of LOCC operations is too restrictive,
and a unique total order and entanglement measure might emerge
when considering more general sets of operations \cite{Audenaert
PE 03}. Both approaches have received some attention but neither
has succeeded completely at the time of writing this article.

In the following we will simply accept the non-uniqueness of
entanglement measures as an expression of the fact that they
correspond to different operational tasks under which different
forms of entanglement may have different degrees of usefulness.

\section{A survey of entanglement measures}

In this section we discuss a variety of bipartite entanglement
measures and monotones that have been proposed in the literature.
All the following quantities are entanglement {\it monotones}, in
that they cannot increase under LOCC. Hence when they can be
calculated they can be used to determine whether certain (finite
or asymptotic) LOCC transformations are possible. However, some
measures have a wider significance that we will discuss as they
are introduced. Before we continue, we consider some features of
the distillable entanglement, particularly with regard to its
computation, as this will be important for some of our later
discussion.

{\it The distillable entanglement --} The distillable entanglement,
$E_D(\rho)$, provides us with the rate at which noisy mixed states
$\rho$ may be converted into the `gold standard' singlet state by
LOCC alone. It formal definition is
\begin{displaymath}
    E_D(\rho) := \sup\left\{ r: \lim_{n\rightarrow\infty}
    \left[\inf_{\Psi} {\rm tr}|\Psi(\rho^{\otimes n})-\Phi(2^{rn})|\right] =
    0 \right\}.
\end{displaymath}
The complexity of this variational definition has the unfortunate
consequence that despite the importance of the distillable
entanglement as an entanglement measure, very little progress has
been made in terms of its computation. It is known for pure states
(where it equals the entropy of entanglement), and for some simple
but very special states \cite{Vedral P 98,Plenio VP 00} (see the end
of this paragraph). To obtain such results and to gain insight into
the amount of distillable entanglement it is particularly important
to be able to provide bounds on its value. {\em Upper bounds} can,
by virtue of eq. (\ref{upperlower}) and requirement 3 for
entanglement monotones, be provided by any other entanglement
monotone and measure but non-monotonic bounds are also of interest
(see the remainder of this section on entanglement measures).
Calculating {\em lower bounds} is more challenging. Some lower
bounds can be obtained by the construction of explicit entanglement
purification procedures \cite{Bennett DSW 96} in particular for Bell
diagonal states \cite{Vollbrecht V 04}. As every state can be
reduced to a Bell diagonal state by random bi-local rotations of the
form $U\otimes U$ (a process known as {\it twirling}), these methods
result in general lower bounds applicable to all states. Improving
these bounds is very difficult as it generally requires the explicit
construction of complex purification procedures in the asymptotic
limit of many copies.

In this context it is of considerable interest to study the {\it
conditional entropy}, which is defined as
$C(A|B):=S(\rho_{AB})-S(\rho_B)$ for a bipartite state
$\rho_{AB}$. It was known for some time that $-C(A|B)$ gives a
lower bound for both the entanglement cost and another important
measure known as the {\it relative entropy of entanglement}
\cite{Plenio VP 00}. However, this bound was also recently shown
to be true for the one way distillable entanglement:
\begin{equation}
    E_D(\rho_{AB}) \geq D_{A\rightarrow B}(\rho_{AB}) \geq
    \max\{S(\rho_B)-S(\rho_{AB}),0\}
\end{equation}
where $D_{A\rightarrow B}$ is the distillable entanglement under the
restriction that the classical communication may only go one way
from Alice to Bob \cite{Devetak W 04}. This bound is known as the
{\it Hashing Inequality} \cite{Bennett DSW 96}, and is significant
as it is a computable, non-trivial, {\it lower bound} to
$E_D(\rho)$, and hence supplies a non-trivial lower bound to many
other entanglement measures. While this bound is generally not
tight, it should be noted that there are examples for which it
equals the distillable entanglement, these include Bell diagonal
states of rank $2$ \cite{Rains 97} and some other special classes of
state such as $\sigma=A|00\rangle\langle 00|+B|00\rangle\langle
11|+B^* |11\rangle\langle 00| +(1-A)|11\rangle\langle 11|$ for which
relative entropy of entanglement (ie an upper bound to $E_D$) can be
computed  \cite{Vedral P 98,Rains 99} and is found to equal the
hashing inequality.

The following subsection will present a variety of other
entanglement measures and quantities that provide upper bounds on
the distillable entanglement. \quad

$\bullet$ {\it Entanglement Cost --} For a given state $\rho$ the
entanglement cost quantifies the maximal possible rate $r$ at which
one can convert blocks of {\it 2-qubit} maximally entangled states
into output states that approximate many copies of $\rho$, such that
the approximations become vanishingly small in the limit of large
block sizes. If we denote a general trace preserving LOCC operation
by $\Psi$, and write $\Phi(K)$ for the density operator
corresponding to the maximally entangled state vector in $K$
dimensions, i.e.\ $\Phi(K)=|\psi_K^+\rangle\langle\psi_K^+|$, then
the entanglement cost may be defined as
\begin{displaymath}
    E_C(\rho) = \inf\left\{ r: \lim_{n\rightarrow\infty}
    \left[\inf_{\Psi} tr|(\rho^{\otimes n}-\Psi(\Phi(2^{rn}))|\right] = 0 \right\}
\end{displaymath}
This quantity is again very difficult to compute indeed. It is
known to equal the entropy of entanglement for pure bi-partite
states. It can also be computed for trivial mixed states
$\rho=\sum_i p_i |\psi_i\rangle\langle\psi_i|$ where the states
$|\psi_i\rangle$ may be discriminated locally perfectly without
destroying the states. A simple example is
$|\psi_1\rangle=|00\rangle$ and
$|\psi_2\rangle=(|11\rangle+|22\rangle)/\sqrt{2}$.

Fortunately, a closely related measure of entanglement, namely the
entanglement of formation, provides some hope as it may actually
equal the entanglement cost. Therefore, we move on to discuss its
properties in slightly more detail.

\quad $\bullet$ {\it Entanglement of Formation --} For a mixed
state $\rho$ this measure is defined as
\begin{equation}
     E_F(\rho) := \inf \{\sum_i p_i E(|\psi_i\rangle\langle
    \psi_i|) ~: ~ \rho = \sum_i p_i |\psi_i\rangle\langle
    \psi_i|\}. \nonumber
\end{equation}
Given that this measure represents the minimal possible average
entanglement over all pure state decompositions of $\rho$, where
$E(|\psi\rangle\langle\psi|)= S({\rm
tr}_B\{|\psi\rangle\langle\psi|\})$ is taken as the measure of
entanglement for pure states, it can be expected to be closely
related to the entanglement cost of $\rho$. Note however that the
entanglement cost is an asymptotic quantity concerning
$\rho^{\otimes n}$ in the limit $n\rightarrow\infty$. It is not
self-evident and in fact unproven that the entanglement of
formation accounts for that correctly. Note however, that the {\it
regularised} or {\it asymptotic} version of the entanglement of
formation, which is defined as $$E^{\infty}_F(\rho) := \lim_{n
\rightarrow \infty} {E_F(\rho^{\otimes n}) \over n}$$ can be
proven rigorously to equal the entanglement cost \cite{Hayden HT
01}, i.e.
\begin{equation}
    E^{\infty}_F(\rho) = E_C(\rho).
\end{equation}
Obviously, the computation of either, the entanglement cost or the
asymptotic entanglement of formation, are extraordinarily difficult
tasks. However, there are indications, though no general proof, that
the entanglement of formation is additive, i.e.
$E_F(\rho)=E_F^{\infty}(\rho)=E_C(\rho)$, a result that would
simplify the computation of $E_C(\rho)$ significantly if it could be
proven. Further to some numerical evidence for the correctness of
this property it is also known that the entanglement of formation is
additive for maximally correlated states in $d\times d$ dimensions,
ie states $\rho_{mc}=\sum_{ij} a_{ij} |ii\rangle\langle jj|$
\cite{Horodecki SS 02}. More generally it is a {\it major} open
question in quantum information to decide whether $E_F$ is a fully
additive quantity, i.e. whether
\begin{equation}
    E_F(\rho^{AB} \otimes \sigma^{AB}) = E_F(\rho^{AB}) +
    E_F(\sigma^{AB}).
\end{equation}
\begin{figure}[th]
\centerline{
\includegraphics[width=7.cm]{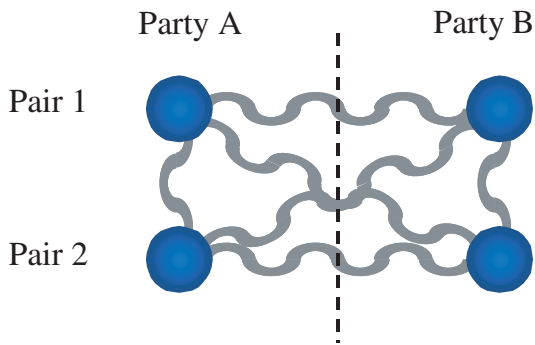}}
\caption{\label{strongsubadd} Schematic picture of the situation
described by eq. (\ref{subadd}). The entanglement of formation of
an arbitrary four particle state $|\psi\rangle$, with particles
held by parties $A$ and $B$ is given is given on the left hand
side of eq. (\ref{subadd}). The right hand side of eq.
(\ref{subadd}) is the sum of the entanglement of formation of the
states $\rho_1=tr_{A_2B_2}|\psi\rangle\langle\psi|$ and
$\rho_2=tr_{A_1B_1}|\psi\rangle\langle\psi|$ obtained by tracing
out the lower  upper half of the system.}
\end{figure}
This problem is known to be equivalent to the {\it strong
superadditivity} of $E_F$
\begin{equation}
    E_F(\rho^{AB}_{12}) ~ ?\geq ? ~ E_F(\rho^{AB}_1) +
    E_F(\rho^{AB}_2)
    \label{subadd}
\end{equation}
where the indices $1$ and $2$ refer to two pairs or entangled
particles while $A$ and $B$ denote the different parties (see fig.
\ref{strongsubadd}).

The importance of these additivity problems is twofold. Firstly,
additivity would imply that $E_F = E_C$ leading to a considerable
simplification of the computation of the entanglement cost.
Secondly, the entanglement of formation is closely related to the
classical capacity of a quantum channel which is given by the Holevo
capacity \cite{Holevo}, and it can be shown that the
additivity of $E_F$ is also {\it equivalent} to the additivity of
the classical communication capacity of quantum channels
\cite{Shor,Audenaert B 04,Pom}!

The variational problem that defines $E_F$ is extremely difficult
to solve in general and at present one must either resort to
numerical techniques for general states \cite{Audenaert VM 01}, or
restrict attention to cases with high symmetry (e.g.
\cite{Vollbrecht W 99,Eisert FPPW 00,Terhal V 00}), or consider
only cases of low dimensionality. Quite remarkably a closed form
solution is known for bi-partite qubit states \cite{Wootters
98,Wootters 01,Audenaert VM 01} that we present here. This exact
formula is based on the often used two-qubit {\em concurrence}
which is defined as
\begin{equation}
    C(\rho) = max\{0,\lambda_1-\lambda_2-\lambda_3-\lambda_4\},
\end{equation}
where the $\lambda_i$ are, in decreasing order, the square roots of
the eigenvalues of the matrix
$\rho\sigma_y\otimes\sigma_y\rho^*\sigma_y\otimes\sigma_y$ where
$\rho^*$ is the elementwise complex conjugate of $\rho$. For general
bi-partite qubit states it has been shown that \cite{Wootters 01}
\begin{equation}
    E_F(\rho) = s(\frac{1+\sqrt{1-C^2(\rho)}}{2})
\end{equation}
with
\begin{equation}
    s(x) = -x\log_2 x - (1-x)\log_2(1-x).
\end{equation}
The two-qubit $E_F(\rho)$ and the two-qubit concurrence are
monotonically related which explains why some authors prefer to
characterise entanglement using only the concurrence rather than the
$E_F$. It should be emphasised however that it is only the
entanglement of formation that is an entanglement measure, and that
the concurrence obtains its meaning via its relation to the
entanglement of formation and not vice versa. For higher dimensional
systems this connection breaks down - in fact there is not even a
unique definition of the concurrence. Therefore, the use of the
entanglement of formation even in the two-qubit setting, is
preferable.

\begin{figure}[th]
\centerline{
\includegraphics[width=7.cm]{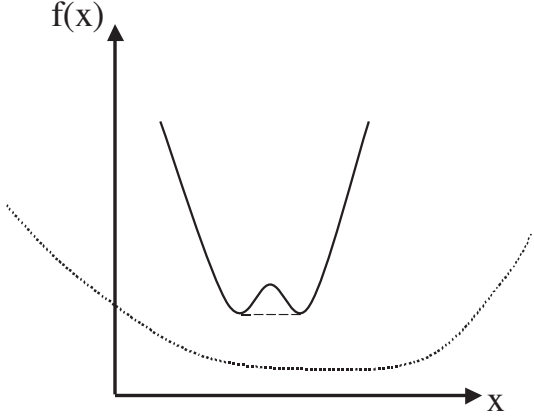}}
\caption{\label{convexroof} A schematic picture of the convex roof
construction in one dimension. The non-convex function $f(x)$ is
given by the solid line. The dotted curve is a convex function
smaller than $f$ and the convex roof, the largest convex function
that is smaller than $f$, is drawn as a dashed curved (it
coincides in large parts with $f$).}
\end{figure}

$\bullet$ {\it Entanglement measures from convex roof constructions
--} The entanglement of formation $E_F$ is an important example of
the general concept of a {\it convex roof} construction. The convex
roof ${\hat f}$ of a function $f$ is defined as the largest convex
function that is for all arguments bounded from above by the
function $f$. A simple example in one variable is given by
$f(x)=x^4-2\alpha^2x^2$ and its convex roof
\begin{displaymath}
    {\hat f}(x)=\left\{ \begin{array}{ll}
        x^4-2\alpha^2x^2 & \mbox{for}\, |x|\ge \alpha\\
        -\alpha^4        &  \mbox{for}\, |x|\le \alpha
    \end{array}\right.
\end{displaymath}
Fig. \ref{convexroof} illustrates this idea graphically with an
example for the convex roof for a function of a single variable.
Generally, for a function $f$ defined on a convex subset of
$\mathbb{R}^n$, the convex roof ${\hat f}$ can be constructed via
the variational problem
\begin{equation}
    {\hat f}(x) = \inf_{x=\sum_{i} p_i x_i} \sum_i p_i f(x_i),
\end{equation}
where the infimum is taken over all possible probability
distributions $p_i$ and choices of $x_i$ such that $x=\sum_{i} p_i
x_i$. It is easy to see that ${\hat f}$ is convex, that ${\hat f}\le
f$ and that any other convex function $g$ that is smaller than $f$
also satisfies $g\le {\hat f}$.

The importance of the convex roof method is based on the fact that
it can be used to construct entanglement monotones from any
unitarily invariant and concave function of density matrices
\cite{Vidal 00}. As this construction is very elegant we will
discuss how it works in some detail. Suppose that we already have
a function $E$ of {\it pure} states, that is known to be an
entanglement monotone on {\it pure} states. This means that for an
LOCC transformation from an input pure state $|\psi\rangle$ to
output pure states $|\psi_i\rangle$ with probability $p_i$, we
have that:
\begin{equation}
    E(|\psi\rangle) \geq \sum_{i} p_i E(|\psi\rangle_i).
\end{equation}
Such pure state entanglement monotones are very well understood, as
it can be shown that a function is a pure state monotone iff it is a
unitarily invariant concave function of the single-site reduced
density matrices \cite{Vidal 00}.

Let us consider the convex-roof extension ${\hat E}$ of such a pure
state monotone $E$ to mixed states.
A general LOCC operation can be written as a sequence of operations
by Alice and Bob. Suppose that Alice goes first, then she will
perform an operation that given outcome $j$ performs the
transformation:
\begin{equation}
\rho \rightarrow \rho_j = {1 \over p_j} \sum_k A_k \rho A_k^\dag
\end{equation}
where the $A_k$ are Alice's local Kraus operators corresponding to
outcome $j$, and {$p_j$=tr$\{\sum_k A_k \rho A_k^\dag\}$} is the
probability of getting outcome $j$. If $k>1$ for any particular
outcome, then Alice's operation is {\it impure}, in that an input
pure state may be taken to a mixed output. However, any such LOCC
{\it impure} operation may be implemented by first performing a LOCC
{\it pure} operation, where Alice and Bob retain information about
all $k$, followed by `forgetting' the values of $k$ at the end
\footnote{i.e. A pure operation is one in which each different
measurement outcome corresponds to only {\it one} Kraus operator}.
It can be shown quite straightforwardly that if an entanglement
measure is convex, then the process of `forgetting' cannot increase
the average output entanglement beyond the average output
entanglement of the intermediate pure operation. Hence if one shows
that a convex quantity is an entanglement monotone for {\it pure}
LOCC operations, then it will be an entanglement monotone in
general.

This means that we need only prove that ${\hat E}$ is an
entanglement monotone for {\it pure} operations acting upon input
mixed states. This can be done as follows \cite{Vidal 00}. Let
$\rho$ be an input state with optimal decomposition $\rho = \sum
q(i) |\phi_i\rangle\langle\phi_i|$, i.e.
\begin{equation}
    {\hat E}(\rho) = \sum_{i} q(i) E(|\phi\rangle_i).
\end{equation}
Suppose that we act upon this state with a measuring LOCC operation,
where outcome $j$ signifies that we have implemented the (not
trace-preserving) {\it pure} map $\Lambda_j$ (i.e. corresponding to
a single Kraus-operator). Let us define:
\begin{eqnarray}
    p(j|i) &&:= \mbox{tr}\{\Lambda_j(|\phi_i\rangle)\},  \nonumber \\
    p(j) &&:= \mbox{tr}\{\Lambda_j(\rho)\}. \nonumber
\end{eqnarray}
It is clear that $p(j)=\sum_i q(i) p(j|i)$, as required by the
standard probabilistic interpretation of ensembles. Hence given
outcome $j$ the state $\rho$ transforms to:
\begin{eqnarray}
    \rho_j &=& {1 \over p(j)} \sum_i q(i) \Lambda_j (|\phi\rangle_i)
    \nonumber \\
    &=& {1 \over p(j)} \sum_i p(i,j) {\Lambda_j (|\phi\rangle_i)\over
    p(j|i)} \nonumber \\
    &=& \sum_i p(i|j) {\Lambda_j (|\phi\rangle_i)\over
    p(j|i)}.
\end{eqnarray}
Hence by the convexity of ${\hat E}$ we have that:
\begin{eqnarray}
    \hat{E} (\rho_j) \leq \sum_i p(i|j) \hat{E} \left( {\Lambda_j (|\phi\rangle_i)\over
    p(j|i)} \right)
\end{eqnarray}
and because ${\hat E}$ is a monotone for operations from pure to
pure states, and as each $\Lambda_j (|\phi\rangle_i)$ is pure by
assertion, we find that:
\begin{eqnarray}
    \sum_j p(j) \hat{E} (\rho_j) &\leq& \sum_j p(j) \sum_i p(i|j) \hat{E} \left( {\Lambda_j (|\phi\rangle_i)\over
    p(j|i)} \right) \nonumber \\
    &=& \sum_i q(i) \sum_j p(j|i) \hat{E} \left( {\Lambda_j (|\phi\rangle_i)\over
    p(j|i)} \right) \nonumber \\
    &\leq& \sum_i q(i) \hat{E} (|\phi_i\rangle) \nonumber \\
    &=& \hat{E}(\rho)
\end{eqnarray}
Hence it can be seen that the convex-roof of any pure state
entanglement monotone is automatically an entanglement monotone for
LOCC transformations from mixed states to mixed states. Together
with the result that a function of pure states is an entanglement
monotone iff it is a unitarily invariant concave function of the
single-site density matrices \cite{Vidal 00}, this provides a very
elegant way of constructing many convex-roof entanglement monotones.
It is interesting to note that although this method can also be used
to construct monotones under separable operations, it does not work
for constructing monotones under the set of PPT transformations, as
unlike the case of LOCC/ separable operations, an {\it impure} PPT
operation {\it cannot} always be equated to a {\it pure PPT}
operation plus {\it forgetting} \cite{Horodecki 01}.

$\bullet$ {\it Relative entropy of entanglement --}  So far we
discussed the extremal entanglement measures, entanglement cost and
entanglement of distillation. For some time it was unclear whether
they were equal or whether there are any entanglement measures that
lie between these two. The regularised version of the {\it relative
entropy of entanglement} provides an example of a measure that lies
between $E_C$ and $E_D$.

One way of understanding the motivation for its definition is by
considering total correlations. These are measured by the quantum
mutual information \cite{Nielsen C 00}
\begin{equation}
    I(\rho_{AB}) = S(\rho_{A}) + S(\rho_{B}) - S(\rho_{AB})\, .
\end{equation}
Employing the {\em quantum relative entropy}
\begin{equation}
    S(\rho||\sigma):= {\rm tr}\{\rho\log\rho-\rho \log{\sigma}\}
\end{equation}
which is a measure of distinguishability between quantum states
one may then rewrite the quantum mutual information as
\begin{equation}
    I(\rho_{AB}) = S(\rho_{AB}||\rho_A\otimes\rho_B)\, .
\end{equation}
If the total correlations are quantified by a comparison of the
state $\rho_{AB}$ with the uncorrelated state
$\rho_A\otimes\rho_B$ then it is intuitive to try and measure the
quantum part of these correlations by a comparison of $\rho_{AB}$
with the closest separable state - a classically correlated state
devoid of quantum correlations. This approach gives rise to the
general definition of the relative entropy of entanglement
\cite{Vedral P 98,Vedral PRK 97,Vedral PJK 97,Plenio VP 00} with
respect to a set $X$ as
\begin{equation}
        E^X_R(\rho):= \inf_{\sigma \in X} S(\rho||\sigma).
        \label{distance}
\end{equation}
This definition leads to a class of entanglement measures known as
the {\it relative entropies of entanglement} (see \cite{Horodecki
HHOSSS 04} for a possible operational interpretation). In the
bipartite setting the set $X$ can be taken as the set of separable
states, states with positive partial transpose, or non-distillable
states, depending upon what you are regarding as `free' states. In
the multiparty setting there are even more possibilities
\cite{Plenio V 01a,Vedral PJK 97} but for each such choice a valid
entanglement measure is obtained as long as the set $X$ is mapped
onto itself under LOCC (one may even consider more general classes
of operations as long as $X$ is mapped onto itself).
\begin{figure}[th]
\centerline{
\includegraphics[width=7.cm]{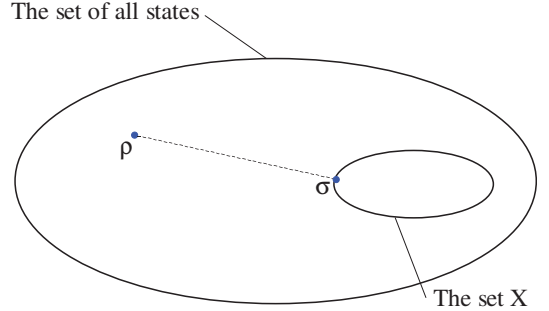}}
\vspace*{0.5cm} \caption{\label{relent} The relative entropy of
entanglement is defined as the smallest relative entropy distance
from the state $\rho$ to states $\sigma$ taken from the set $X$.
The set $X$ may be defined as the set of separable states,
non-distillable states or any other set that is mapped onto itself
by LOCC. }
\end{figure}
Employing the properties of the quantum relative entropy it is
then possible to prove that it is a convex entanglement measure
satisfying all the conditions 1 - 4 \cite{Vedral P 98} which is
also asymptotically continuous \cite{Donald H 99}. The bipartite
relative entropies have been used to compute tight upper bounds to
the distillable entanglement of certain states \cite{Audenaert
EJPVD 01}, and as an invariant to help decide the asymptotic
interconvertibility of multipartite states \cite{Linden PSW
99,Galvao PV 00,Plenio I}. The relative entropy of entanglement is
bounded from below by the conditional entropy
\begin{displaymath}
    E_R(\sigma) \ge \max\{S(\sigma_A),S(\sigma_B)\}-S(\sigma_{AB})
\end{displaymath}
which can be obtained from the fact that for any bi-partite
non-distillable state $\rho$ we have
\begin{eqnarray*}
    S(\sigma_A)+S(\sigma_A||\rho_A) &\le&
    S(\sigma_{AB})+S(\sigma_{AB}||\rho_{AB}),\\
    S(\sigma_B)+S(\sigma_B||\rho_B) &\le&
    S(\sigma_{AB})+S(\sigma_{AB}||\rho_{AB}).
\end{eqnarray*}
The relative entropy measures are generally not additive, as
bipartite states can be found where
\begin{equation}
    E^{X}_R(\rho^{\otimes n}) \neq ~ n E^X_R(\rho).
\end{equation}
The regularized relative entropy of entanglement
\begin{displaymath}
   E^{\infty}_{R \, , X}:= \lim_{n\rightarrow\infty} \frac{E^X_R(\rho^{\otimes n})}{n}
\end{displaymath}
is therefore of some interest. In various cases exhibiting a high
degree of symmetry the regularised versions of some relative entropy
measures can be calculated employing ideas from semi-definite
programming and optimization theory \cite{Boyd V 04}. These cases
include the {\it Werner states}, i.e. states that are invariant
under the action of unitaries of the form $U\otimes U$, and which
take the form $\sigma(p) = p\sigma_a + (1-p)\sigma_s$, where $p\in
(1/2,1]$ and $\sigma_a$ ($\sigma_s$) are states proportional to the
projectors onto the anti-symmetric (symmetric) subspace. It can be
shown that \cite{Audenaert EJPVD 01}
\begin{eqnarray}
    E^{\infty}_{R \, , PPT}(\sigma(p)) =
        \left\{ \begin{array}{ll}
    1-H(p), & p\leq \frac{d+2}{2d} \\
        & \\
    \lg\frac{d+2}{d}+(1-p)\lg\frac{d-2}{d+2}, & p>\frac{d+2}{2d}
    \end{array} \right.\label{theorem}
\end{eqnarray}
where $H(p) = -p\lg p-(1-p)\lg(1-p)$. It is notable that while
this expression is continuous in $p$ it is not differentiable for
$p=1/2+1/d$. These results can be extended to the more general
class of states that is invariant under the action of $O\otimes
O$, where $O$ is an orthogonal transformation \cite{Audenaert
DVW02}.

{\em Other distance based measures --} In eq. (\ref{distance}) one
may consider replacing the quantum relative entropy by different
distance measures to quantify how far a particular state is from a
chosen set of disentangled states. Many interesting examples of
other functions that can be used for this purpose may be found in
the literature (see e.g. \cite{Vedral PRK 97,Vedral P 98,Wei G 03}).
It is also worth noting that the relative entropy functional is {\it
asymmetric}, in that $S(\rho||\sigma) \neq S(\sigma||\rho)$. This is
connected with asymmetries that can occur in the discrimination of
probability distributions \cite{Vedral P 98}. One can consider
reversing the arguments and tentatively define an LOCC monotone
$J^X(\rho):=\inf \{S(\sigma||\rho)\,:\, \sigma \in X\}$. The
resulting function has the advantage of being additive, but
unfortunately it has the problem that it can be infinite on pure
states \cite{Eisert AP 03}. An additive measure that does not suffer
from this deficiency will be presented later on in the form of the
`squashed' entanglement.

 $\bullet$ {\it The Distillable Secret Key--} The
Distillable Secret Key, $K_D(\rho)$, quantifies the asymptotic
rate at which Alice and Bob may distill secret classical bits from
many copies of a shared quantum state. Alice and Bob may use a
shared quantum state to distribute a classical bit of information
- for instance if they share a state $1/2(|00\rangle\langle
00|+|11\rangle\langle 11|)$, then they may measure it in the
$|0\rangle,|1\rangle$ basis to obtain an identical classical bit
$0,1$, which could form the basis of a cryptographic protocol such
as one-time pad (see e.g. \cite{Nielsen C 00} for a description of
one-time pad). However,if we think of a given bipartite mixed
state $\rho_{AB}$ as the reduction of a pure state held between
Alice, Bob, and a malicious third party Eve, then it is possible
that Eve could obtain information about the secret bit from
measurements on her subsystem. In defining $K_D$ it is assumed
that each copy of $\rho_{AB}$ is purified {\it independently} of
the other copies. If we reconsider the example of the state
$1/2(|00\rangle\langle 00|+|11\rangle\langle 11|)$, we can easily
see that it is not secure. For instance, it could actually be a
reduction of a GHZ state $|000\rangle + |111\rangle$ held between
Alice, Bob and Eve, in which case Eve could also have complete
information about the `secret' bit. The quantity $K_D$ is hence
zero for this state, and is in fact zero for all separable states.

One way of getting around the problem of Eve is to use entanglement
distillation. If Alice and Bob distill bipartite pure states, then
because pure states must be uncorrelated with any environment, any
measurements on those pure states will be uncorrelated with Eve.
Moreover, if the distilled pure states are EPR pairs, then because
each local outcome $|0\rangle,|1\rangle$ occurs with equal
probability, each EPR pair may be used to distribute exactly 1
secret bit of information. This means that $K_D(\rho) \geq D(\rho)$.
However, entanglement distillation is not the only means by which a
secret key can be distributed, it examples of PPT states are known
where $K_D(\rho) > 0$, even though $D(\rho)=0$ for all PPT states
\cite{Horodecki HHO 05}. It has also been shown that the regularized
relative entropy with respect to separable states is an upper bound
to the distillable secret key, $E^{\infty}_{R \, , SEP}(\rho) \geq
K_D(\rho)$ \cite{Horodecki HHO 05}.

$\bullet$ {\it Logarithmic Negativity --} The partial transposition
with respect to party $B$ of a bipartite state $\rho_{AB}$ expanded
in a given local orthonormal basis as $\rho = \sum
\rho_{ij,kl}|i\rangle\langle j|\otimes |k\rangle\langle l|$ is
defined as
\begin{equation}
    \rho^{T_B} := \sum_{i,j,k,l}\rho_{ij,kl} |i\rangle\langle j| \otimes |l\rangle \langle
    k|.
\end{equation}
The spectrum of the partial transposition of a density matrix is
independent of the choice of local basis, and is independent of
whether the partial transposition is taken over party $A$ or party
$B$. The positivity of the partial transpose of a state is a
necessary condition for separability, and is sufficient to prove
that $E_D(\rho)=0$ for a given state \cite{Peres,Horodecki
97,Horodecki HH 98}. The quantity known as the {\it Negativity}
\cite{Zyczkowski HSL 98,Eisert P 99}, $N(\rho)$, is an
entanglement monotone \cite{Lee KPL 00,PhD,Vidal W 02,Plenio 05}
that attempts to quantify the negativity in the spectrum of the
partial transpose. We will define the Negativity as
\begin{equation}
        N(\rho):= \frac{||\rho^{T_B}||-1}{2},
\end{equation}
where $||X||:=$tr$\sqrt{X^{\dag}X}$ is the trace norm. While being a
convex entanglement monotone, the negativity suffers the deficiency
that it is not additive. A more suitable choice for an entanglement
monotone may therefore be the so called {\it Logarithmic Negativity}
which is defined as
\begin{equation}
        E_N(\rho):= \log_2 ||\rho^{T_B}||\, .
\end{equation}
The monotonicity of the negativity immediately implies that $E_N$
is an entanglement monotone that cannot increase under the more
restrictive class of deterministic LOCC operations, ie
$\Phi(\rho)=\sum_i A_i\rho A_i^{\dagger}$. While this is not
sufficient to qualify as an entanglement monotone it can also be
proven that it is a monotone under probabilistic LOCC
transformations \cite{Plenio 05}. It is additive by construction
but fails to be convex. Although $E_N$ is manifestly continuous,
it is not asymptotically continuous, and hence does not reduce to
the entropy of entanglement on all pure states.

The major practical advantage of $E_N$ is that it can be
calculated very easily. In addition it also has various
operational interpretations as an upper bound to $E_D(\rho)$, a
bound on teleportation capacity \cite{Vidal W 02}, and an
asymptotic entanglement cost for exact preparation under the set
of PPT operations \cite{Audenaert PE 03}.

$\bullet$ {\em The Rains bound --} The logarithmic negativity,
$E_N$, can also been combined with a relative entropy concept to
give another monotone known as the {\it Rains' Bound} \cite{Rains
01}, which is defined as
\begin{equation}
        B(\rho) := \min_{all~states~\sigma} \left[ S(\rho || \sigma) +
        E_N(\sigma)\right]\, .
\end{equation}
The function $S(\rho || \sigma) + E_N(\sigma)$ that is to be
minimized is not convex which suggests the existence of local
minima making the numerical minimization infeasible. Nevertheless,
this quantity is of considerable interest as one can observe
immediately that $B(\rho)$ is a lower bound to $E^{PPT}_R(\rho)$
as $E_N(\sigma)$ vanishes for states $\sigma$ that have a positive
partial transpose. It can also be shown that $B(\rho)$ is an upper
bound to the Distillable Entanglement. It is interesting to
observe that for Werner states $B(\rho)$ happens to be equal to
$\lim_{n \rightarrow \infty}E^{PPT}_R(\rho^{\otimes n})/n$
\cite{Audenaert EJPVD 01,Rains 01}, a connection that has been
explored in more detail in \cite{Audenaert DVW02,Audenaert PE
03,Ishizaka}.

$\bullet$ {\it Squashed entanglement --} Another interesting
entanglement measure is the squashed entanglement \cite{Christandl
W 03} (see also \cite{Tucci}) which is defined as
\begin{eqnarray}
    &~& E_{sq} := \inf \left[ {1 \over 2} I(\rho_{ABE}) ~~:~ {\rm tr}_E\{\rho_{ABE}\}=\rho_{AB} \right] \nonumber\\
    &~& {\rm where: } \nonumber \\
    &~& I(\rho_{ABE}) := S(\rho_{AE})+S(\rho_{BE})-S(\rho_{ABE})-S(\rho_{E})\, . \nonumber
\end{eqnarray}
In this definition $I(\rho_{ABE})$ is the {\it quantum conditional
mutual information}, which is often also denoted as $I(A;B|E)$. The
motivation behind $E_{sq}$ comes from related quantities in
classical cryptography that determine correlations between two
communicating parties and an eavesdropper. The squashed entanglement
is a convex entanglement monotone that is a lower bound to
$E_F(\rho)$ and an upper bound to $E_D(\rho)$, and is hence
automatically equal to $S(\rho_A)$ on pure states. It is also
additive on tensor products, and is hence a useful non-trivial lower
bound to $E_C(\rho)$. It has furthermore been proven that the
squashed entanglement is continuous \cite{Alicki F 03}, which is a
non-trivial statement because in principle the minimization must be
carried out over {\em all} possible extensions, including infinite
dimensional ones. Note that despite the complexity of the
minimization task one may find upper bounds on the squashed
entanglement from explicit guesses which can be surprisingly sharp.
For the totally anti-symmetric state $\sigma_a$ for two qutrits one
obtains immediately (see Example 9 in \cite{Christandl W 03}) that
$E_D(\sigma_a)\le E_{sq}(\sigma_a) \le \log_2\sqrt{3}$ which is very
close to the sharpest known upper bound on the distillable
entanglement for this state which is $\log_2 5/3$ \cite{Rains
01,Audenaert EJPVD 01}. The Squashed entanglement is also known to
be lockable \cite{Christandl W 03,Christandl 06}, and is an upper
bound to the secret distillable key \cite{Christandl 06}.

$\bullet$ {\it Robustness quantities and norm based monotones --}
This paragraph discusses various other approaches to entanglement
measures and then moves on to demonstrate that they and some of
the measures discussed previously can actually be placed on the
same footing.

{\em Robustness of Entanglement --} Another approach to
quantifying entanglement is to ask how much noise must be mixed in
with a particular quantum state before it becomes separable. For
example
\begin{equation}
P(\rho):=\inf_\sigma \{ \lambda \,|\, \sigma \,\mbox{a state}\,;\,(1
- \lambda) \rho + \lambda \sigma \, \in \,SEP \,;\, \lambda \geq 0
\} \label{lam}
\end{equation}
measures the minimal amount of {\it global} state $\sigma$ that must
be mixed in to make $\rho$ separable. Despite the intuitive
significance of equation (\ref{lam}), for mathematical reasons it is
more convenient to parameterize this noise in a different way:
\begin{eqnarray}
R_g(\rho) :=&& \inf t \nonumber \\
\mbox{such that } && t \geq 0 \nonumber \\
\mbox{and } &&  \exists \mbox{ a state } \sigma \nonumber \\
\mbox{ such that } &&  \rho + t \sigma \mbox{ is separable.}
\nonumber
\end{eqnarray}
This quantity, $R_g$, is known as the {\it Global Robustness} of
entanglement \cite{Harrow N}, and is monotonically related to
$P(\rho)$ by the identity $P(\rho) = R_g(\rho)/(1+R_g(\rho))$.
However, the advantage of using $R_g(\rho)$ rather than $P(\rho)$
is that the first quantity has very natural mathematical
properties that we shall shortly discuss. The global robustness
mixes in arbitrary noise $\sigma$ to reach a separable state,
however, one can also consider noise of different forms, leading
to other forms of robustness quantity. For instance the earliest
such quantity to be defined, which is simply called the {\it
Robustness}, $R_s$, is defined exactly as $R_g$ except that the
noise $\sigma$ must be drawn from the set of separable states
\cite{Vidal T 99,Steiner 03,Brandao 05b}. One can also replace the
set of separable states in the above definitions with the set of
PPT states, or the set of non-distillable states. The robustness
monotones can often be calculated or at least bounded
non-trivially, and have found applications in areas such as
bounding fault tolerance \cite{Harrow N, Virmani HP 05}.

{\em Best separable approximation --} Rather than mixing in
quantum states to destroy entanglement one may also consider the
question of how much of a separable state is contained in an
entangled state. The ensuing monotone is known as the {\it Best
Separable Approximation} \cite{Lewenstein S 98}, which we define
as
\begin{eqnarray}
    BSA(\rho) :=&& \inf \mbox{tr} \{\rho - A \}
    \nonumber \\ \mbox{ such that } &&  A \geq 0 \,\,;\,\,A \in \,SEP
    \nonumber \\ \mbox{ and } &&  (\rho - A) \geq 0 \nonumber.
\end{eqnarray}
This measure is not easy to compute analytically or numerically.
Note however, that replacing the set SEP by the set PPT allows us
to write this problem as a semidefinite programme \cite{Boyd V 04}
for which efficient algorithms are known.

{\em One shape fits all --} It turns out that the robustness
quantities, the best separable approximation as well as the
negativity are all part of a general family of entanglement
monotones. Such connections were first observed in \cite{Vidal W
02}, where it was noted that the Negativity and Robustness are part
of a general family of monotones that can be constructed via a
concept known as a {\it base norm} \cite{Hartkamper}. We will
explain this connection in the following. However, our discussion
will deviate a little from the arguments presented in \cite{Vidal W
02}, as this will allow us to include a wider family of entanglement
monotones such as $R_g(\rho)$ and $BSA(\rho)$.

To construct this family of monotones we require two sets $X,Y$ of
operators satisfying the following conditions: (a) $X,Y$ are
closed under LOCC operations (even measuring ones), (b) $X,Y$ are
convex cones (i.e. also closed under multiplication by
non-negative scalars), (c) each member of $X\,(Y)$ can be written
in the form {$\alpha_{X(Y)}\times$positive-semidefinite operator},
where $\alpha_{X\,(Y)}$ are fixed real constants, and (d) any
Hermitian operator $h$ may be expanded as:
\begin{equation}
    h = a \Omega - b \Delta
\end{equation}
where $\Omega \in X, \Delta \in Y$ are normalised to have trace
$\alpha_X,\alpha_Y$ respectively, and $a,b \geq 0$. Given two such
sets $X,Y$ and any state $\rho$ we may define an entanglement
monotone as follows:
\begin{eqnarray}
R_{X,Y}(\rho):= \inf_{\Omega \in X, \Delta \in Y} \{b ~~~ | ~\rho =
a \Omega - b \Delta, a,b \geq 0\} \label{rdef}
\end{eqnarray}
Note that if $\Omega, \Delta$ are also constrained to be quantum
states (i.e. $\alpha_X=\alpha_Y=1$), then we may rewrite this
equation:
\begin{eqnarray}
    &~& R_{X,Y}(\rho) = \nonumber \\
    &~&\inf \{b ~|~b \geq 0,~ \exists \Delta \in Y,\Omega \in X \mbox{ s.t. }{\rho + b \Delta \over 1 + b} = \Omega \}
    \nonumber
\end{eqnarray}
Hence equation (\ref{rdef}) defines a whole family of quantities
that have a similar structure to robustness quantities.

In the more general case where $\alpha_X,\alpha_Y\neq 1$, the
quantities $R_{X,Y}(\rho)$ will not be robustness measures, but they
will still be entanglement monotones. This can be shown as follows,
where we will suppress the subscripts $_{X,Y}$ for clarity. Suppose
that a LOCC operation acts on $\rho$ to give output $\rho_i =
\Lambda_i(\rho)/q_i$ with probability $q_i$. Suppose also that the
optimum expansion of the initial state $\rho$ is:
\begin{eqnarray}
\rho = a \Omega - R \Delta \nonumber
\end{eqnarray}
Then the output ensemble can be written as:
\begin{eqnarray}
&& \{ q_i ~ ; ~  {a \Lambda_i(\Omega) - R \Lambda_i(\Delta) \over
q_i}
\} \nonumber \\
\equiv && \{ q_i ~ ; ~ \tilde{a}_i { \alpha_X \Lambda_i(\Omega)
\over \mbox{tr}\{\Lambda_i(\Omega)\}}  - \tilde{R}_i {\alpha_Y
\Lambda_i(\Delta) \over \mbox{tr}\{\Lambda_i(\Delta)\}}  \}
\label{expansion}
\end{eqnarray}
where
\begin{eqnarray}
\tilde{a}_i = {a \, \mbox{tr} \{ \Lambda_i (\Omega)\} \over \alpha_X
q_i} \,\,\,\,\,\,\,\,\,  ;  \,\,\,\,\,\,\,\, \tilde{R}_i = {R \,
\mbox{tr} \{ \Lambda_i (\Delta) \} \over \alpha_Y q_i} \nonumber
\end{eqnarray}
Now because of the structure of each operator in $X,Y$, we have that
$\tilde{a}_i,\tilde{R}_i \geq 0$, and hence for each outcome $i$ the
expansion in (\ref{expansion}) is a valid decomposition. This means
that the average output entanglement satisfies:
\begin{eqnarray}
\sum_i q_i R(\rho_i) \leq \sum_i q_i \tilde{R}_i = R \sum_i
{\mbox{tr} \{ \Lambda_i (\Delta) \} \over \alpha_Y } = R
\end{eqnarray}
and hence the $R_{X,Y}$ give entanglement monotones. It is also not
difficult to show that the $R_{X,Y}$ are convex functions. In the
case that the two sets $X$ and $Y$ are identical, then the quantity
\begin{equation}
||h||_{X,X}:= \inf_{\Omega, \Delta \in X} \{a + b~|~ h = a \Omega -
    b\Delta,
    a,b \geq 0\}\, . \nonumber
\end{equation}
can be shown to be a norm, and in fact it is a norm of the so-called
{\it base norm} kind. As $||h||_{X,X}$ can be written as a simple
function of the corresponding $R_{X,X}$, this gives the robustness
quantities a further interesting mathematical structure.

All the monotones mentioned at the beginning of this subsection
fit into this family - the `{\it Robustness}' arises when both
$X,Y$ are the set of separable operators; the `{\it Best Separable
approximation}' arises when $X$ is the set of separable operators,
$Y$ is the set {$\{$positive semi-definite operators$\times-1\}$};
the global robustness arises when $X$ is the set of separable
operators, $Y$ is the set of all positive semidefinite operators
\cite{Vidal T 99,Steiner 03,Brandao 05b,Harrow N}; the Negativity
arises when $||\rho||_{X,Y}$ where both $X,Y$ are the set of
normalised Hermitian matrices with positive partial transposition.
Note that the `{\it Random Robustness}' is not a monotone and so
does not fit into this scheme, for definition and proof of
non-monotonicity see \cite{Vidal T 99,Steiner 03}.

{\em The greatest cross norm monotone --} Another form of norm
based entanglement monotone is the {\it cross norm} monotone
proposed in \cite{Rudolph,Rudolph 01,Rudolph 02}. The {\it
greatest cross norm} of an operator $A$ is defined as:
\begin{equation}
||A||_{\rm{gcn}} := \inf \left[ \sum_{i=1}^{n} ||u_i||_1 ||v_i||_1
~:~ A = \sum_i u_i \otimes v_i \right] \label{rud}
\end{equation}
where $||y||_1 := {\rm tr}\{\sqrt{y^{\dag}y}\}$ is the trace norm,
and the infimum is taken over all decompositions of $A$ into finite
sums of product operators. For finite dimensions it can be shown
that a density matrix $\rho_{AB}$ is separable iff
$||\rho||_{\rm{gcn}}$=1, and that the quantity:
\begin{equation}
E_{\rm{gcn}}(\rho) := ||\rho||_{\rm{gcn}} -1
\end{equation}
is an entanglement monotone \cite{Rudolph,Rudolph 01,Rudolph 02}.
As it is expressed as a complicated variational expression,
$E_{\rm{gcn}}(\rho)$ can be difficult to calculate. However, for
pure states and cases of high symmetry it may often be computed
exactly. Although $E_{\rm{gcn}}(\rho)$ does not fit precisely into
the family of base norm monotones discussed above, there is a
relationship. If the sum in (\ref{rud}) is restricted to {\it
Hermitian} $u_i$ and $v_i$ (which is of course only allowed if $A$
is Hermitian), then we recover precisely the base norm
$||A||_{X,Y}$, where $X,Y$ are taken as the set of separable
states. Hence $E_{\rm{gcn}}$ is an upper bound to the robustness
\cite{Rudolph,Rudolph 01,Rudolph 02}.

$\bullet$ {\it Entanglement Witness monotones --} Entanglement
Witnesses are tools used to try to determine whether a state is
separable or not. A Hermitian operator $W$ is defined as an
Entanglement Witness if:
\begin{eqnarray}
    \forall \,\, \rho \in SEP \!\!\!\!\!\!&&\!\!\!\!\!\! \mbox{tr}\{W \rho \} \geq 0\nonumber\\
    &and& \\
    \exists \rho \,\,\,\mbox{s.t.}\!\!\!\!\!\!\!&& \!\!\!\!\!\!\!\mbox{tr}\{W \rho \} <
    0.
    \nonumber
\end{eqnarray}
Hence $W$ acts as a linear hyperplane separating some entangled
states from the convex set of separable ones.
\begin{figure}[th]
\centerline{
\includegraphics[width=7.cm]{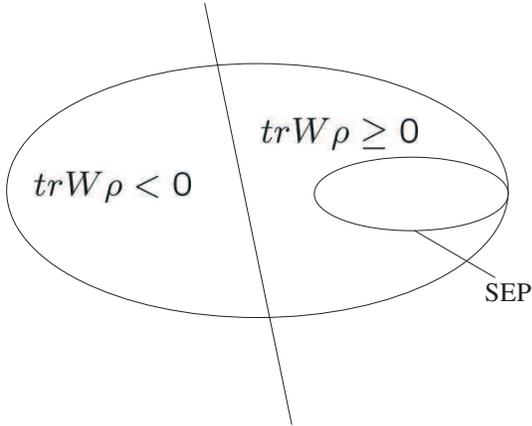}}
\caption{\label{Witness} An entanglement witness is a Hermitean
operator defining a hyperplane in the space of positive operators
such that for all separable states we have $tr W\rho \ge 0$ and
there is a $\rho$ for which $tr W\rho <0$.  }
\end{figure}
Many entanglement witnesses are known, and in fact the CHSH
inequalities are well known examples. One can take a suitable
Entanglement Witness (EW) and use the amount of `violation'
\begin{equation}
    E_{wit}(W) = \max \{ 0, -\mbox{tr}\{W \rho \}\}
\end{equation}
as a measure of the non-separability of a given state. Many
entanglement monotones can be constructed by choosing (bounded) sets
of of EWs and defining monotones as the minimal violation over all
witnesses taken from the chosen set - see e.g. \cite{Brandao 05b}.
It turns out that this approach also offers another unified way of
understanding the robustness and negativity measures discussed in
the previous item \cite{Brandao 05b}.

\medskip

This concludes our short survey of basic entanglement measures. Our
review has mostly been formulated for two-party systems with finite
dimensional constituents. In the remaining two subsections we will
briefly summarize the problems that we are faced with in more
general settings - where we are faced with more parties and infinite
dimensional systems. We will present some of the results that have
been obtained so far, and highlight some unanswered questions.\\

\medskip

\section{Infinite dimensional systems} In the preceding sections
we have explicitly considered only finite dimensional systems.
However, one may also develop a theory of entanglement for the
infinite dimensional setting. This setting is often also referred
to as the {\it continuous variable} regime, as infinite
dimensional pure states are usually considered as wavefunctions in
continuous position or momentum variables. The quantum harmonic
oscillator is an important example of a physical system that needs
to be described in an infinite dimensional Hilbert space, as it is
realized in many experimental settings, e.g. as modes of quantized
light.

{\em General states --} A naive approach to infinite dimensional
systems encounters several complications, in particular with regards
to continuity. Firstly, we will need to make some minimal
requirements on the Hilbert space, namely that the system has the
property that {tr$\{\exp^{H/T}<\infty\}$} to avoid pathological
behaviour due to limit points in the spectrum \cite{footnote3}. The
harmonic oscillator is an example of a system satisfying this
constraint. Even so, without further constraints, entanglement
measures cannot be continuous because by direct construction one may
demonstrate that in any arbitrarily small neighborhood of a pure
product state, there exist pure states with {\it arbitrarily strong}
entanglement as measured by the entropy of entanglement \cite{Eisert
SP 02}. The following example makes this explicit. Chose
$\sigma_0=|\psi_0\rangle\langle\psi_0|$ where $|\psi_0\rangle =
|\phi_{A}^{(0)}\rangle\otimes |\phi_{B}^{(0)}\rangle$, and consider
a sequence of pure states $\sigma_k=|\psi_k\rangle\langle\psi_k|$
defined by
\begin{eqnarray}
    |\psi_k\rangle &=&
    \sqrt{1-\epsilon_k}
    |\psi_0\rangle + \sqrt{\frac{\epsilon_k}{k}}
    \sum_{n=1}^k |\phi_{A}^{(n)}\rangle\otimes |\phi_{B}^{(n)}\rangle,
\end{eqnarray}
where $\epsilon_k = 1/ \log (k)^2$ and
$\{|\phi_{A/B}^{(n)}\rangle: n\in{\mathbb{N}}_0\}$ are orthonormal
bases. Then $\{\sigma_k\}_{k=1}^{\infty}$ converges to $\sigma_0$
in trace-norm, i.e., $\lim_{k\rightarrow \infty} \| \sigma_k -
\sigma_0\|_1=0$ while $\lim_{k\rightarrow
\infty}E(\sigma_k)=\infty$. Obviously, $E$ is not continuous
around the state $\sigma_0$.

However, this perhaps surprising feature can only occur if the
mean energy of the states $\sigma_k$ grows unlimited in $k$. If
one imposes additional constraints such as restricting attention
to states with bounded mean energy then one finds that the
continuity of entanglement measures can be recovered \cite{Eisert
SP 02}. More precisely, given the Hamiltonian $H$ and the set
${\cal S}_M = \{\rho\in{\cal S}|tr[\rho H]\le M\}$ where ${\cal
S}$ is the set of all density matrices, then we find for example
that for $\sigma\in {\cal S}_M({\cal H})$, $M>0$, being a pure
state that is supported on a finite-dimensional subspace of ${\cal
S}({\cal H})$, and $\{\sigma_n\}_{n=1}^{\infty}$, $\sigma_n\in
{\cal S}_{n M}({\cal H}^{\otimes n})$, being a sequence of states
satisfying
\begin{equation}
    \lim_{n\rightarrow\infty}
    \|\sigma_n -\sigma^{\otimes n}\|=0,
\end{equation}
then
\begin{equation}
    \lim_{n\rightarrow\infty}
    \frac{|E_F(\sigma^{\otimes n})
    - E_F(\sigma_n)|}{n}
    =0.\label{Asy}
\end{equation}
Similar statements hold true for the entropy of entanglement and the
relative entropy of entanglement. The technical details can be found
in \cite{Eisert SP 02}. Even with this constraint however, the
description of entanglement and its quantification is
extraordinarily difficult, although some concrete statements can be
made \cite{Parker BP 00}. Note however, that for continuous
entanglement measures that are strongly super-additive (in the sense
of eq. (\ref{subadd}) in the situation given in fig.
\ref{strongsubadd}) one can provide lower bounds on entanglement
measures in terms of a simpler class of state, the Gaussian states
\cite{Wolf GC 05}. This motivates the consideration of more
constrained sets of states.

{\bf\em Gaussian states --} A further simplification that can be
made is to consider only the set of {\it Gaussian} quantum states.
This set of states is important because not only do they play a
key role in several fields of theoretical and experimental
physics, but they also have some attractive mathematical features
that enable many interesting problems to be tackled using basic
tools from linear algebra. We will concentrate on this class of
states, as they have been subject to the most progress. The
systems that are being considered possess $n$ canonical degrees of
freedom representing for example $n$ harmonic oscillators, or $n$
field modes of light. These canonical operators are usually
arranged in vector form
\begin{equation}
    {{O}}=(O_1,\ldots,O_{2n})^T=
    (X_1,P_1,\ldots,X_{n},P_{n})^T.
\end{equation}
Then the canonical commutation relations take the form $[O_j,O_k]=i
\sigma_{j,k}$, where we define the {\it symplectic} matrix as
follows:
\begin{equation}
    \sigma :=\bigoplus_{j=1}^{n}
    \left[\begin{array}{rr}
    0 & 1 \\
    -1 & 0 \\
    \end{array}
    \right].
\end{equation}
States $\rho$ may now also be characterized by functions that are
defined on phase space. Given a vector $\xi\in{\mathbb{R}}^{2n}$,
the Weyl or Glauber operator is defined as:
\begin{equation}
    W_{\xi} = e^{i  \xi^{T} \sigma O}.
\end{equation}
These operators generate displacements in phase space, and are used
to define the {\it characteristic function} of $\rho$:
\begin{equation}
    \chi_{\rho}(\xi) = \mbox{tr}[\rho W_{\xi}].
\end{equation}
This can be inverted by the transformation \cite{Cahill}:
\begin{equation}
    \rho = \frac{1}{(2\pi)^n} \int d^{2n}\xi
    \chi_{\rho}(-\xi) W_{\xi},
\end{equation}
and hence the characteristic function uniquely specifies the state.
Gaussian states are now defined as those states whose characteristic
function is a Gaussian \cite{Eisert P 03}, i.e.,
\begin{equation}
    \chi_{\rho}(\xi) =
    \chi_{\rho}(0) e^{-\frac{1}{4}\xi^T \Gamma \xi + D^T
    \xi},
\end{equation}
where $\Gamma$ is a $2n\times 2n$-matrix and $D\in{\mathbb{R}}^{2n}$
is a vector. In defining Gaussian states in this way it is easy to
see that the reduced density matrix of any Gaussian state is also
Gaussian - to compute the characteristic function of a reduced
density matrix we simply set to zero any components of $\xi$
corresponding to the modes being traced out.

As a consequence of the above definition, a Gaussian
characteristic function can be characterized via its first and
second moments only, such that a Gaussian state of $n$ modes
requires only $2n^2+n$ real parameters for its full description,
which is polynomial rather than exponential in $n$. The first
moments form the displacement vector $ d_j= \langle
O_j\rangle_\rho= \mbox{tr}[O_j \rho ]$ $j=1,...,2n$ which is
linked to the above $D$ by $D= \sigma d$. They can be made zero by
means of a unitary translation in the phase space of individual
oscillators and carry no information about the entanglement
properties of the state.

The second moments of a quantum state are defined as the expectation
values $\langle O_j O_k \rangle$. Because of the canonical
commutation relationships the value of $\langle O_k O_j \rangle$ is
fixed by the value of $\langle O_j O_k \rangle$ (the operators
$O_j$,$O_k$ either commute, or their commutator is proportional to
the identity), and so all second moments can be embodied in the real
symmetric $2n\times 2n$ covariance matrix $\gamma$ which is defined
as
\begin{eqnarray}
    \gamma_{j,k}
    &=&
    2 \mbox{Re} \, \mbox{tr}\left[
    \rho \left(O_j-\langle O_j\rangle_{\rho} \right)
    \left(O_k-\langle O_k\rangle_{\rho} \right)
    \right] \nonumber\\
    &=& \mbox{tr}\left[
    \rho \left( \{O_j,O_k\} - 2 \langle O_j\rangle_{\rho} \langle O_k\rangle_{\rho} \right)
    \right]
\end{eqnarray}
where $\{\}$ denotes the anticommutator. The link to the above
matrix $\Gamma$ is $\Gamma= \sigma^{T}\gamma \sigma$. With this
convention, the covariance matrix of the $n$-mode vacuum is simply
$\id_{2n}$. Clearly, not all real symmetric $2n\times 2n$-matrix
represent quantum states as these must obey the Heisenberg
uncertainty relation. In terms of the second moments the
`uncertainty principle' can be written as the matrix inequality
\begin{equation}\label{res}
    \gamma + i \sigma\geq 0 \, .
\end{equation}
Note that for one mode this uncertainty principle is actually {\it
stronger} than the usual Heisenberg uncertainty principle
presented in textbooks, and in fact equation (\ref{res}) is the
strongest uncertainty relationship that may be imposed on the
2nd-moments $\langle O_j O_k \rangle$. This is because it turns
out that {\it any} real symmetric matrix $\gamma$ satisfying the
uncertainty principle (\ref{res}) corresponds to a valid quantum
state. Proving equation (\ref{res}) is actually not too difficult
\cite{uncert,Simon MD 94} - we start with a $2n$ component vector
of complex numbers $y$, and define an operator $Y:=\sum_j y_j (O_j
- \langle O_j \rangle)$. Then the positivity of $\rho$ implies
that tr$\{\rho Y^\dag Y\} \geq 0 ~ \forall ~ y$. A little algebra,
and use of the canonical commutation relationships, shows that
tr$\{\rho Y^\dag Y\} \geq 0 ~ \forall ~ y ~ \Leftrightarrow ~
\gamma + i \sigma\geq 0$.

This observation has quite significant implications concerning the
separability of two-mode Gaussian states shared by two parties.
Indeed, a necessary condition for the separability of Gaussian
states can be formulated on the basis of the partial transposition,
or more precisely partial time reversal, expressed on the level of
covariance matrices. In a system with canonical degrees of freedom
time reversal is characterized by the transformation that leaves the
positions invariant but reverses the relevant momenta $X \mapsto X,
\; P \mapsto -P$. A two-party Gaussian state is then separable
exactly if the covariance matrix corresponding to the partially
transposed state again satisfies the uncertainty relations
\cite{Simon 00,Duan GCZ 00,Werner W 01,Giedke KLC 01,Giedke DCZ 01}.
More advanced questions concerning the interconvertibility of pairs
of states under local operations can also often be answered fully in
terms of the elements of the covariance matrix \cite{Eisert P
02,Eisert SP 02a,Giedke C 02,Giedke ECP 03}. In particular, the
question of the interconvertability of pure bi-partite Gaussian
states of an arbitrary number of modes can be decided in full
generality \cite{Giedke ECP 03}.

{\em Gaussian operations --} The development of the theory of
entanglement of Gaussian states requires also the definition of
the concept of Gaussian operations. Gaussian operations may be
defined as those operations that map {\em all} Gaussian input
states onto a Gaussian output state. This definition is not
constructive but fortunately more useful characterizations exist.
Physically useful is the fact that Gaussian operations correspond
exactly to those operations that can be implemented by means of
optical elements such as beam splitters, phase shifts and
squeezers together with homodyne measurements \cite{Giedke C
02,Eisert SP 02a,Fiuraszek 02}.

The most general real linear transformation S which implements the
mapping
\begin{equation}
    S:O\longmapsto O'=S O
\end{equation}
will have to preserve the canonical commutation relations
$[O'_j,O'_k]=i\sigma_{jk}\id$ which is exactly the case if $S$
satisfies
\begin{equation}
    S\sigma S^T=\sigma \, .
\end{equation}
This condition is satisfied by the real $2n\times 2n$ matrices $S$
that form the so-called real symplectic group $Sp(2n,\mathbb{R})$.
Its elements are called symplectic or canonical transformations.
It is useful to know that any orthogonal transformation is
symplectic. To any symplectic transformation $S$ also
$S^T,S^{-1},-S$ are symplectic. The inverse of $S$ is given by
$S^{-1}= \sigma S^T \sigma^{-1}$ and the determinant of every
symplectic matrix is $\det[ S] =1$ \cite{Arvind DMS 95,Simon SM
87}.
Given a real symplectic transformation $S$ there exists a unique
unitary transformation $U_S$ acting on the state space such that
the Weyl operators satisfy $U_S W_{\xi} U_S^\dagger = W_{S\xi} $
for all $\xi\in {\mathbb{R}}^{2}$. On the level of covariance
matrices $\gamma$ of an $n$-mode system a symplectic
transformation $S$ is reflected by a congruence
\begin{equation}
    \gamma \longmapsto S \gamma S^T .
\end{equation}
Generalized Gaussian quantum operations may also be defined
analogously to the finite dimensional setting, ie by appending
Gaussian state ancillas, performing joint Gaussian unitary
evolution followed by tracing out the ancillas or performing
homodyne detection on them \cite{Giedke C 02,Eisert SP
02a,Fiuraszek 02,Eisert P 03}.

{\em Normal forms --} Given a group of transformations on a set of
matrices it is always of great importance to identify normal forms
for matrices that can be achieved under this group of
transformations. Of further interest and importance are invariants
under the group transformations. For the set of Hermitean matrices
and the full unitary group these correspond to the concepts of
diagonalization and eigenvalues. In the setting of covariance
matrices and the symplectic group we are led to the Williamson
normal forms and the concept of symplectic eigenvalues. Indeed,
Williamson \cite{Williamson 36} (see \cite{Simon MD 94} for a more
easily accessible reference) proved that for any covariance matrix
$\Gamma$ on $n$ harmonic oscillators there exists a symplectic
transformation $S$ such that
\begin{equation}
    S\Gamma S^T = \bigoplus_{j=1}^n\left(\begin{array}{cc} \mu_j & 0 \\ 0 & \mu_j
    \end{array}\right)
\end{equation}
The diagonal elements $\mu_i$ are the so-called {\em symplectic
eigenvalues} of a covariance matrix $\Gamma$ which are the
invariants under the action of the symplectic group. The set
$\{\mu_1,\ldots,\mu_n\}$ is usually referred to as the {\em
symplectic spectrum}. The symplectic spectrum can be obtained
directly from the absolute values of the eigenvalues of
$i\sigma^{-1}\Gamma$. The transformation to the Williamson normal
form implements a normal mode decomposition thereby reducing any
computational problem, such as the computation of the entropy, to
that for individual uncoupled modes. Each block in the Williamson
normal form represents a thermal state for which the evaluation of
most physical quantities is straightforward.

{\bf\em Entanglement quantification --} Equipped with these tools
we may now proceed to discuss the quantification of entanglement
in the Gaussian continuous variable arena. Despite all the above
technical tools the quantification of entanglement for Gaussian
states is complicated and only very few measures may be defined
let alone computed.

$\bullet$ {\em Entropy of entanglement:} On the level of pure
state we may again employ the entropy of entanglement which we may
now express in terms of the covariance matrix. Assume Alice and
Bob are in possession of $n_A+n_B$ harmonic oscillators in a
Gaussian state described by the covariance matrix $\Gamma$ and
Alice holds $n_A$ of these oscillators. Then it can be shown that
the entropy of entanglement is given by
\begin{equation}
    S = \sum_{i=1}^{n_A}\left( \frac{\mu_i+1}{2}\log_2\frac{\mu_i+1}{2}
    -\frac{\mu_i-1}{2}\log_2\frac{\mu_i-1}{2}\right)
\end{equation}
where the $\mu_i$ are the {\it symplectic} eigenvalues of Alice's
reduced state described by the covariance matrix $\Gamma_A$ which is
simply the submatrix of $\Gamma$ referring to the system pertaining
to Alice. These symplectic eigenvalues are, as remarked above, the
positive eigenvalues of $i\sigma^{-1}\Gamma_A$. The proof of the
above formula is obtained by transforming the covariance to its
Williamson normal form and subsequently determine the entropy of the
single mode states. Note that on the set of Gaussian states the
entropy is evidently continuous and it can be shown that this
remains the case for the set of states with bounded mean energy
\cite{Eisert SP 02}.

$\bullet$ {\em Entanglement of formation:} In the finite
dimensional setting the defintion of the entanglement of formation
has been unambiguous. In the Gaussian state setting however this
is no longer the case. One may define the entanglement of
formation of a Gaussian state either (i) with respect to
decompositions in pure Gaussian states or (ii) with respect to
decompositions in arbitrary pure states. In case (i) it has been
proven that the so-defined entanglement of formation is an
entanglement monotone under Gaussian operations and that it can be
computed explicitly in the case where both parties hold a single
harmonic oscillator each. Remarkably, this entanglement of
formation is even additive for symmetric two-mode states
\cite{Wolf GKWC 03}. For the case of a single copy of a mixed
symmetric Gaussian two mode state it can also be demonstrated that
the definition (i) coincides with definition (ii) \cite{Giedke
WKWC 03,Wolf GKWC 03}. The entanglement of formation can be shown
to be continuous for systems with energy constraint \cite{Shirokov
04}.

$\bullet$ {\em Distillable Entanglement:} The distillable
entanglement in the continuous variable setting is, as expected,
extremely difficult to compute. Furthermore, its definition is not
unambiguous as one may define distillation with respect to (i)
Gaussian operations only, or (ii) general quantum operations. It is
remarkable that it has been proven that the setting (i) does not
actually permit entanglement distillation at all \cite{Eisert SP
02a,Giedke C 02,counter}. Therefore, non-Gaussian operations need to
be considered. Then, in setting (ii), for Gaussian states it can be
shown to be continuous and interestingly it can also be demonstrated
that for any $\rho$ there exists a Gaussian state $\rho_G$ with the
same first and second moments such that $E_D(\rho_G)\le D(\rho)$.
Finding explicit procedures implementing distillation protocols is
very difficult which makes it very difficult to determine lower
bounds on the distillable entanglement. Various other measures of
entanglement, such as those described below, may be used to find
upper bounds on the distillable entanglement.

$\bullet$ {\em Relative entropy of entanglement:} As for the
entanglement of formation there are now various possible
definitions of the relative entropy of entanglement all of which
are at least as difficult to compute as in the finite dimensional
setting. If the relative entropy of entanglement should serve as a
provable upper bound on the distillable entanglement under general
LOCC, then it will have to be computed with respect to the set of
separable general continuous variable states. This is obviously a
very involved quantity and only known on pure states where it
equals the entropy of entanglement. If one considers the relative
entropy of entanglement of a state with bounded mean energy with
respect to the unrestricted set of separable states, then it can
be shown that the relative entropy of entanglement is continuous
\cite{Eisert SP 02}. A more tractable setting is that of the
relative entropy of entanglement with respect to the set of
Gaussian separable states but in this case its interpretation is
unclear.

$\bullet$ {\em Logarithmic negativity:} As in the finite
dimensional setting, most entanglement measures are exceedingly
difficult to compute. The exception is again the logarithmic
negativity which is an entanglement monotone \cite{Plenio 05} but
differs, on pure states, from the entropy of entanglement. For a
system of $n=n_A+n_B$ harmonic oscillators in a Gaussian state
described by the covariance matrix $\Gamma$, the logarithmic
negativity can again be expressed in terms of symplectic
eigenvalues. Indeed, considering the covariance matrix
$\Gamma^{T_B}$ of the partially transposed state we find
\begin{equation}
    E_N = -\sum_{i=1}^{n} \log_2[\min(1,{\tilde\mu}_k)]
\end{equation}
where the ${\tilde\mu}_k$ form the symplectic spectrum for the
partially transposed state described by covariance matrix
$\Gamma^{T_B}$, ie the symplectic eigenvalues. This formula is
again proven by applying a normal mode decomposition, this time to
the partially transposed covariance matrix, reducing the problem
to a single mode question. It is interesting to note that on
Gaussian states the logarithmic negativity also possesses an
interpretation as a special type of entanglement cost
\cite{Audenaert PE 03}.

The tools for the manipulation and quantification are used in the
assessment of the quality of practical optical entanglement
manipulation protocols. It should also be noted that these tools
have been used successfully to study entanglement properties of
quasi-free fields on lattices (i.e. lattices of harmonic
oscillators) initiating the study of the scaling behaviour of
entanglement between contiguous blocks in the ground state of
interacting quantum systems \cite{Audenaert EPW 02}. The above
methods and quantities permitted the rigorous proofs of the scaling
of entanglement between contiguous blocks in the ground state of a
linear harmonic chain with a Hamiltonian that is quadratic in
position and momentum\cite{Audenaert EPW 02} and a rigorous
connection between the entanglement of an arbitrary set of harmonic
oscillators and its surrounding with the boundary area \cite{Plenio
EDC 05,Cramer EPD 06}. This illustrates the usefulness of the
results that have been obtained in continuous variable entanglement
theory over the last years.

\medskip

{\bf Multi-particle entanglement --} Although the two-party
setting has provided many interesting examples of quantum
entanglement, the multiparty setting allows us to explore a much
wider range of effects. Phenomena such as quantum computation
especially when based on cluster states \cite{Raussendorf B 01},
entanglement enhanced measurements \cite{Wineland BIMH 92,Huelga
MPEPC 97}, multi-user quantum communication \cite{Murao JPV
98,Hillery BB 99,Karlsson KI 99,Murao PV 00} and the GHZ paradox
all require consideration of systems with more than two particles.
For this reason it is important to investigate entanglement in the
multi-party setting. We will proceed along similar lines to the
bi-partite setting, first discussing briefly basic properties of
states and operations and then describing various approaches to
the quantification of multi-particle entanglement.

{\em States and Operations --} In the following we are going to
concentrate again on local operations and classical communication
whose definition extend straightforwardly to the multi-party
setting. Some remarks will also be made concerning PPT operations
which are here defined as operations that preserve ppt-ness of
states across {\em all} possible bi-partite splits. That is, any
three-party state shared between $A$, $B$ and $C$ that remains
positive under partial transposition of particle $A$ or $B$ or $C$
is mapped again onto a state with this property.

In the bi-partite setting we initiated our discussions with the
identification of some general properties of multi-party entangled
states such as the identification of disentangled states and
maximally entangled states. At this stage crucial differences
between the two-party and the multi-party setting become apparent.
Let us begin by trying to identify the equivalent of the two-party
maximally entangled states. In the bi-partite setting we already
identified qubit states of the form
$(|00\rangle+|11\rangle)/\sqrt{2}$ maximally entangled because
every other qubit state can be obtained from it with certainty
using LOCC only. One natural choice for a state with this property
could be the GHZ-state
\begin{equation}
    |GHZ\rangle = \frac{1}{\sqrt{2}}(|0\rangle_A|0\rangle_B|0\rangle_C
    + |1\rangle_A|1\rangle_B|1\rangle_C).
\end{equation}
This state has the appealing property that its entanglement across
any bi-partite cut e.g. party A versus parties B and C assume the
largest possible value of 1 ebit. Also, a local measurement in the
$|\pm\rangle=(|0\rangle\pm|1\rangle)/\sqrt{2}$ basis for example
on party A allows us to create deterministically a maximally
entangled two-party state of parties $B$ and $C$. Then we can
obtain any other two-party entangled state for parties $B$ and $C$
by LOCC. Unfortunately, however there are tri-partite entangled
states that cannot be obtained from the GHZ state using LOCC
alone. One such example is the W-state \cite{Dur W state}
\begin{equation}
    |W\rangle = \frac{1}{\sqrt{3}}(|0\rangle_A|0\rangle_B|1\rangle_C
    + |0\rangle_A|1\rangle_B|0\rangle_C + |1\rangle_A|0\rangle_B|0\rangle_C).
\end{equation}
Note however that LOCC operations applied to a GHZ-state allow us
to approximate the W-state as closely as we like, albeit with
decreasing success probability. In the four party setting however
it can be shown that there are pairs of pure states that cannot
even be obtained from each other approximately employing LOCC
alone \cite{Verstraete DVdeM 02}. This clearly shows that on the
single-copy level it is not possible to establish a generic notion
of a maximally entangled state.

Of course we have already learnt in the bi-partite setting that
the requirement of exact transformations on single copies can lead
to phenomena such as incomparable states and does not yield a
simple and unified picture of entanglement. In the bi-partite
setting such a unified picture for pure state entanglement emerges
however in the asymptotic setting of arbitrarily many identically
prepared states. One might therefore wonder whether a similar
approach will be successful in multi-partite systems. These hopes
will be dashed in the following. In the asymptotic setting we
would need to establish the possibility for the reversible
interconversion in the asymtptotic setting. If that were possible
we could rightfully claim that all tri-partite entanglement is
essentially equivalent and only appears in different
concentrations that we could then quantify unambiguously. The
simplest situation that one may consider to explore this
possibility is the interconversion between GHZ and the EPR pairs
across parties $AB$, $AC$ and $BC$, ie in the limit
$N\rightarrow\infty$ we would like to see
\begin{equation}
    |GHZ\rangle^{\otimes N} \leftrightharpoons
    |EPR\rangle_{AB}^{\otimes n_{AB}}\otimes
    |EPR\rangle_{AB}^{\otimes n_{AC}}\otimes
    |EPR\rangle_{AB}^{\otimes n_{BC}}\, .
\end{equation}
To decide this question one needs to identify sufficiently many
entanglement monotones. In the case of reversibility these
entanglement monotones will remain constant. The local entropies
represent such a monotone. These are not enough to decide the
question but it turns out that $E_{R}(\rho_{AB})+S(\rho_{AB})$, ie
the sume of the relative entropy of entanglement of the reduction
to two parties and the entanglement between these two parties and
the third, is also an entanglement monotone in this setting. This
is then sufficient to prove that the above process cannot be
achieved reversibly \cite{Linden PSW 99}. This result suggests
that as opposed to the bi-partite setting there is not such a
simple and unique concept of a maximally entangled state in the
multi-partite setting.

One may however try and and make progress by generalizing the idea
of a single entangled state from which all other states can be
obtained reversibly in the asymptotic setting. Instead one may
consider a set of states from which all other state may be obtained
asymptotically reversibly. The smallest such set is usually referred
to as an MREGS which stands for Minimal Reversible Entanglement
Generating Set \cite{Bennett MREGS}. It was natural to try and see
whether the set $\{|GHZ\rangle_{ABC}, |EPR\rangle_{AB},
|EPR\rangle_{AC}, |EPR\rangle_{BC}\}$ is sufficient to generate the
W-state reversibly. Unfortunately, even this conjecture was proved
wrong \cite{Acin multiparty,Galvao PV 00}. Similar results have also
been obtained in the four-party setting \cite{Wu Z 00}. Therefore,
an MREGS would also have to contain the W-state as well. It is
currently an open question whether under LOCC operations any finite
MREGS actually exists.

In another approach to overcome the difficulties presented above
one may consider extensions of the set of operations that is
available for entanglement transformations. A natural
generalization are PPT operations that have already made an
appearance in the bi-partite setting. Adopting PPT operations
indeed simplifies the situation somewhat. In the single copy
setting any k-partite entangled state can be transformed, with
finite success probability, into any other k-partite entangled
state by PPT operations \cite{Ishizaka 04,Ishizaka P 05}. The
success probabilities can be surprisingly large, e.g. the
transformation from GHZ to W state succeeds with more than $75\%$
\cite{Ishizaka 04}. It is noteworthy that PPT operations also
overcome the constraint that is imposed by the non-increase of the
Schmidt-number under LOCC. Indeed, PPT operations (and also the
use of LOCC with bound entanglement as a free resource) allow us
to increase the Schmidt number. This result already implicit in
\cite{Audenaert PE 03} was made explicit in \cite{Audenaert PE
03,Ishizaka 04}. It was hoped for that this strong increase in
probabilities and the vanishing of the Schmidt number constraint
would lead to reversibility in the multi-partite setting, ie a
finite MREGS under PPT operations. This question is however still
remains open \cite{Plenio I}.

Up until now we have restricted attention to pure multi-party
entangled states. Now let us consider the definition of separable
multi-particle states. The most natural definition for
disentangled states arises from the idea that we call a state
disentangled if we can create it from a pure product state by the
action of LOCC only. This implies that separable states are of the
form
\begin{equation}
    \rho = \sum_i p_i \rho^i_A \otimes \rho^i_B \otimes \rho^i_C
    \otimes...
\end{equation}
where the $A,B,C..$ label different parties. However, one can go
beyond this definition. Indeed, the state
$(|00\rangle_{AB}+|11\rangle_{AB})/\sqrt{2}\otimes |0\rangle_C$ is
clearly entangled and therefore not separable in the above sense.
However, it also does not exhibit three-party entanglement as the
third party $C$ is uncorrelated from the other two. Therefore may
call this tri-partite state 2-entangled. One may now try and
generalize this idea to mixed states. For example we could define
as the set of 2-entangled states any $\rho$ that may be written in
the form
\begin{equation}
    \rho = \sum_i p_i \rho_A^{(i)}\otimes \rho_{BC}^{(i)} +
    \sum_i q_i \rho_B^{(i)}\otimes \rho_{AC}^{(i)} +
    \sum_i r_i \rho_C^{(i)}\otimes \rho_{AB}^{(i)}
\end{equation}
with positive $p_i,q_i$ and $r_i$. Then, for $N$ parties one may
then define k-entangled states as a natural generalization of the
above 3-party definition. While this definition appears natural it
encounters problems when we consider several identical copies of
states of the form given above. In that case one can obtain a
3-entangled state by LOCC acting on two copies of the above
2-entangled state. As a simple example consider a three party
state where Alice has two qubits and Bob and Charlie each hold
one. Then a state of the form: ${1 \over 2}[ |0\rangle\langle
0|_{A1} \otimes EPR(A2,B) \otimes |0\rangle\langle 0|_{C} +
|1\rangle\langle 1|_{A1} \otimes |0\rangle\langle 0|_{B} \otimes
EPR(A2,C)]$ is only 2-entangled. However, given two copies of this
state the `classical flag' particle $A1$ can enable Alice to
obtain (with some probability) one EPR pair with Bob, and one with
Charlie. She can then use these EPR pairs and teleportation to
distribute any three party entangled state she chooses. States of
three qubits displaying a similar phenomenon can also be
constructed. Hence we are faced with a subtle dilemma - either
this notion of `k-entanglement' is not closed under LOCC, or it is
not closed under taking many copies of states. Note however that
these states may still have relevance for example in the study of
fault-tolerant quantum computation \cite{Virmani HP 05}.

{\em Quantifying Multi-partite entanglement --} Already in the
bi-partite setting it was realized that there are many
non-equivalent ways to quantify entanglement \cite{Virmani P 00}.
This concerned mainly the mixed state case, while in the pure
state case the entropy of entanglement is a distinguished measure
of entanglement. In the multipartite setting this situation
changes. As was discussed above it appears difficult to establish
a common currency of multipartite entanglement even for pure
states due to the lack of asymptotically reversible
interconversion of quantum states. The possibility to define
k-entangled states and the ensuing ambiguities lead to additional
difficulties in the definition of entanglement measures in
multi-partite systems.

Owing to this there are many ways to go about quantifying
multipartite entanglement. Some of these measures will be natural
generalizations from the bi-partite setting while others will be
specific to the multi-partite setting. These measures and their
known properties will be the subject of the remainder of this
section.

{\em Entanglement Cost and Distillable Entanglement --} In the
bi-partite setting it was possible to define unambiguously the
entanglement of pure states establishing a common ''currency'' for
entanglement. This then formed the basis for unique definitions of
the entanglement cost and the distillable entanglement. The
distillable entanglement determined the largest rate, in the
asymptotic limit, at which one may obtain pure maximally entangled
states from an initial supply of mixed entangled states using LOCC
only. However, in the multi-particle setting there is no unique
target state that one may aim for. One may of course provide a
target state specific definition of distillable entanglement, for
example the largest rate at which one may prepare GHZ states
\cite{Murao PPVK 98}, cluster states \cite{Aschauer DB 04,Chen L
04} or any other class that one is interested in. As these
individual resources are not asymptotically equivalent each of
these measures will capture different properties of the state in
question.

One encounters similar problems when attempting to define the
entanglement cost. Again, one may use singlet states as the
resource from which to construct the state by LOCC but one may
also consider other resources such as GHZ or W states. For each of
these settings one may then ask for the best rate at which one can
create a target state using LOCC in the asymptotic limit.
Therefore we obtain a variety of possible definitions of
entanglement costs.

While the interpretation of each of these measures is clear it is
equally evident that it is not possible to arrive at a unique
picture from abstract considerations alone. The operational point
of view becomes much more important as different resources may be
readily available in different experimental settings and then
motivating different definitions of the entanglement cost and the
distillable entanglement.

$\bullet$ {\it Relative Entropic Measures. Distance measures --}
In the bipartite setting we have discussed various distance based
measures in which one minimizes the distance of a state with
respect to a set of states that does not increase in size under
LOCC. One such set was that of separable states and a particularly
important distant measure is the relative entropy of entanglement.
This lead to the relative entropy of entanglement. As we discussed
in the first part of this section the most natural extension of
the definition of separable states in the multipartite setting is
given by
\begin{equation}
    \rho = \sum_i p_i \rho^i_A \otimes \rho^i_B \otimes \rho^i_C
    \otimes...
\end{equation}
where the $A,B,C..$ label different parties. In analogy with the
bipartite definition one can hence define a multipartite relative
entropy measure:
\begin{equation}
    E^X_R(\rho):= \inf_{\sigma \in X} S(\rho||\sigma)
\end{equation}
where $X$ is now the set of multipartite separable states. As in
the bipartite case the resulting quantity is an entanglement
monotone which, for pure states, coincides with the entropy of
entanglement. Therefore, on pure states, this measure is additive
while it is known to be sub-additive on mixed states. Remarkably,
the multipartite relative entropy of entanglement is {\it not}
even additive for pure states - a counterexample is provided by
the totally anti-symmetric state
\begin{equation}
    |A\rangle = \frac{1}{\sqrt{6}}\sum_{ijk}
    \epsilon_{ijk}|ijk\rangle
\end{equation}
where $\epsilon_{ijk}$ is the totally anti-symmetric tensor
\cite{Plenio I}. One can also compute the relative entropy of
entanglement for some other tri-partite states. Examples of
particular importance in this respect are the W-state for which we
find
\begin{equation}
    E_R{|W\rangle} = \log_2\frac{9}{4}
\end{equation}
and the states $|GHZ(\alpha)\rangle=\alpha|000\rangle + \beta
|111\rangle$ for which we find
\begin{equation}
    E_R{|W\rangle} = -|\alpha|^2 \log_2 |\alpha|^2 -
    |\beta|^2\log_2 |\beta|^2\, .
\end{equation}
More examples can be found quite easily.

Also in our discussion of multi-partite entanglement we introduced
the notion of k-entangled states. let us denote the set of
k-entangled state of an N-partite system by ${\cal S}_k^{N}$. If
ew explicitly consider the single copy setting, then it is clear
that that the set ${\cal S}_k^{N}$ does not increase under LOCC.
As a consequence it can be used as the basis for generalizations
of the relative entropy of entanglement simply replacing the set
$X$ above by ${\cal S}_k^{N}$. We have learnt however that the set
${\cal S}_k^{N}$ may grow when allowing for two or more copies of
the state. This immediately implies that the so constructed
measure will exhibit sub-additivity again. Given that even the
standard definition for the multi-partite relative entropy of
entanglement is sub-additive this should not be regarded as a
deficiency. Indeed, this subadditivity may be viewed as a strength
as it could lead to particularly strong bounds on the associated
distillable entanglement.

Exactly the same principle may be used to extend any of the
distance based entanglement quantifiers to multi-party systems -
one simply picks a suitable definition of the `unentangled' set
$X$ (i.e. a set which is closed under LOCC operations, and
complies with some notion of locality), and then defines the
minimal distance from this set as the entanglement measure. As
stated earlier, one may also replace the class of separable states
with other classes of limited entanglement - e.g. states
containing only bipartite entanglement. Such classes are {\it not}
in general closed under LOCC in the many copy setting and so the
resulting quantities may exhibit strong subadditivity and their
entanglement monotonicity needs to be verified carefully.

$\bullet$ {\it Robustness measures. Norm based measures.} The
robustness measures discussed in the bipartite case extend
straightforwardly to the multiparty case. In the bipartite case we
constructed the robustness monotones from two sets of operators
$X,Y$ that were closed under LOCC operations, and in addition
satisfied certain convexity and `basis' properties. To define
analogous monotones in the multiparty case we must choose sets of
multiparty operators that have these properties. One could for
example choose the sets $X,Y$ to be the set of k-separable
positive operators, for any integer $k$.

$\bullet$ {\it Entanglement of Assistance. Localizable
entanglement. Collaborative Localizable entanglement.} One way of
characterizing the entanglement present in a multiparty state is
to understand how local actions by the parties may generate
entanglement between two distinguished parties. For example, in a
GHZ state $1/\sqrt{2}(|000\rangle + |111\rangle)$ of three
parties, it is possible to generate an EPR pair between any two
parties using only LOCC operations - if one party measures in the
$1/\sqrt{2}(|0\rangle \pm |1\rangle)$ basis, then there will be a
residual EPR pair between the remaining two parties. This is the
case even though the reduced state of the two parties is by itself
unentangled. The first attempt to quantify this phenomenon was the
{\it Entanglement of Assistance} proposed by \cite{DiVincenzo
FMSTU}. The {\it Entanglement of Assistance} is a property of
3-party states, and quantifies the maximal bipartite entanglement
that can be generated on average between two parties $A,B$ if
party C measures her particle and communicates the result to
$A,B$. A related measure known as the {\it Localizable
Entanglement} was proposed and investigated in \cite{Verstraete DC
04,Verstraete PC 04,Pachos P 04,Popp VMC 05} for the general
multiparty case - this is defined as the maximum entanglement that
can be generated between two parties if all {\it remaining} $n$
parties act using LOCC on the particles that they possess
\cite{Smolin VW 05}. Both these measures require an underlying
measure of bipartite entanglement to quantify the the entanglement
between the two singled-out parties. In the original articles
\cite{DiVincenzo FMSTU, Verstraete DC 04} the pure state entropy
of entanglement was used, however, one can envisage the use of
other entanglement measures \cite{Gour S 05}. The Localizable
Entanglement has been shown to have interesting relations to
correlation functions in condensed matter systems \cite{Verstraete
DC 04,Verstraete PC 04,Pachos P 04,Popp VMC 05}.

As multiparty entanglement quantifiers, both the Entanglement of
Assistance and the Localizable entanglement have the drawback that
they can deterministically {\it increase} under LOCC operations
between all parties \cite{Gour S 05}. This phenomenon occurs because
these measures are defined under the restriction that Alice and Bob
cannot be involved in classical communication with any other parties
- it turns out that in some situations allowing this communication
can increase the entanglement that can be obtained between Alice and
Bob \cite{Gour S 05}. This observation lead the authors of
\cite{Gour S 05} to define the {\it Collaborative} Localizable
Entanglement as the maximal bipartite entanglement (according to
some chosen measure) that may be obtained (on average) between Alice
and Bob using LOCC operations involving {\it all} parties. It is
clear that by definition these collaborative entanglement measures
are entanglement monotones.

It is interesting to note that although the bare Localizable
entanglement is not a monotone, its regularised version {\it is} a
monotone for multiparty pure states \cite{Horodecki OW 05}. In
\cite{Horodecki OW 05} it is shown that the regularised version of
the Localizable entanglement reduces to the minimal entropy of
entanglement across any bipartite cut that divides Alice and Bob,
which is clearly a LOCC monotonous quantity by the previous
discussion of bipartite entanglement measures.

$\bullet$ {\it Geometric measure.} In the case of pure multiparty
states one could try to quantify the `distance' from the set of
separable states by considering various functions of the maximal
overlap with a product state \cite{Wei G 03}. One interesting
choice of function is the logarithm. This was used in \cite{Wei
EGM 03} to define the following entanglement quantifier:
\begin{equation}
G(|\psi\rangle) := - \log \left\{ \sup (|\langle \psi | \alpha
\otimes \beta \otimes \gamma ...\rangle|^2) \right\},
\end{equation}
where the supremum is taken over all pure product states. This
quantity is non-negative, equals zero iff the state $|\psi\rangle$
is separable, and is manifestly invariant under local unitaries. One
can extend this quantity to mixed states using a convex roof
construction. However $G$ is not an entanglement monotone, and it is
{\it not} additive for multiparty pure states \cite{Werner H 02}.
Nevertheless, $G$ is worthy of investigation as it has useful
connections to other entanglement measures, and also has an
interesting relationship with the question of channel capacity
additivity \cite{Werner H 02}. We could also have described $G$ as a
norm based measure, as the quantity $\sup (|\langle \psi | \alpha
\otimes \beta \otimes \gamma ...\rangle|)$ is a norm (of vectors)
known to mathematicians as the {\it injective tensor norm}
\cite{Floret}.

$\bullet$ {\it `Tangles' and related quantities. Entanglement
quantification by local invariants.} An interesting property of
bipartite entanglement is that it tends to be {\it monogamous}, in
the sense that if three parties $A,B,C$ have  the same dimensions,
and if two of the parties $A$ and $B$ are very entangled, then a
third party $C$ can only be weakly entangled with either $A$ or $B$.
 If $AB$ are in a singlet state then they cannot be entangled
with C at all. In \cite{Coffman KW 00}  this idea was put into the
form of a rigorous inequality  for three qubit states using a
entanglement quantifier known as the {\it tangle}, $\tau (\rho)$.
 For a $qubit \times n$ dimensional systems the tangle is
defined as
\begin{equation}
\tau (\rho) = \left\{ \inf \sum_i p_i C^2(| \psi_i \rangle\langle
\psi_i| ) \right\}
\end{equation}
where $C^2(| \psi \rangle\langle \psi|)$ is the square of the
concurrence of pure state $|\psi\rangle$ and the infimum is taken
over all pure state decompositions. The concurrence can be used in
this way as any pure state of a $2\times n$ system is equivalent
to a two qubit pure state. It has been shown that $\tau(\rho)$
satisfies the inequality \cite{Coffman KW 00,Osborne V 05}
\begin{equation}
    \tau (A:B) + \tau (A:C) + \tau (A:D) + ... \leq \tau(A:BCD...)
\nonumber
\end{equation}
where the notation $A:X_1X_2...$ means that $\tau$ is computed
across the bipartite splitting between party $A$ and parties
$X_1X_2..$. This shows that the amount of bipartite entanglement
between party $A$ and several individual parties $B,C,D,..$ is
bounded from above by the amount of bipartite entanglement between
party $A$ and parties $BCD...$ collectively.

In the case of three qubit pure states the {\it residual tangle}
\begin{equation}
    \tau_3 = \tau(A:BC) - \tau (A:B) - \tau (A:C) \nonumber
\end{equation}
is a local-unitary invariant that is independent of which qubit is
selected as  party  $A$, and might be proposed as a `quantifier' of
three party entanglement for pure states of 3-qubits. However, there
are states with genuine three party entanglement for which the
residual tangle can be zero (the W-state serves as an example
\cite{Coffman KW 00}). However, the residual tangle can only be
non-zero if there is genuine tripartite entanglement, and hence can
be used as a indicator of three party entanglement.

Another way to construct multiparty entanglement measures for
multi-qubit {\it pure} systems is simply to single out one qubit,
compute the entanglement between that qubit and the rest of the
system, and then average over all possible choices of the singled
out qubit. As any {\it pure} bipartite system of dimensions
$2\times m$ can be written in terms of two Schmidt coefficients,
one can apply all the formalism of two-qubit entanglement. This
approach has been taken, for example, in the paper by Meyer and
Wallach \cite{Meyer W 01}. That the quantity proposed in
\cite{Meyer W 01} is essentially only a measure of the bipartite
entanglement across various splittings was shown by Brennen
\cite{Brennen 03}. Extensions of this approach are presented in
\cite{Emary 04}.

{\em Local unitary invariants:} The residual tangle is only one of
many {\it local} unitary invariants that have been developed for
multiparty systems. Such local invariants are very important for
understanding the structure of entanglement, and have also been
used to construct prototype entanglement measures. Examples of
local invariants that we have already mentioned are the Schmidt
coefficients and the Geometric measure. In the multiparty case we
may define the {\it local} invariants as those functions that are
invariant under a {\it local} group transformation of fixed
dimensions. If each particle is assumed for simplicity to have the
same dimension $d$, then these local groups are of the form $A
\otimes B \otimes C ...$ where $A,B,C..$ are taken from a
particular $d$-dimensional group representation such as the
unitary group $U(d)$ or the group of invertible matrices $GL(d)$.
The physical significance of the local $GL(d)$ invariants is that
if two states have different values for such an invariant then
they cannot even be inter-converted probabilistically using
stochastic LOCC (`SLOCC') operations. In the case of local unitary
groups one typically only need consider invariants that are {\it
polynomial} functions of the density matrix elements - this is
because it can be shown that two states are related by a local
unitary iff they have the same values on the set of polynomial
invariants \cite{braunschweig}. For more general groups a complete
set of polynomial invariants cannot always be constructed, and one
must also consider local invariants that are not polynomial
functions of states - one example is a local $GL$ invariant called
the `Schmidt rank', which is the minimal number of product
state-vector terms in which a given multiparty pure state may be
coherently expanded. It can be shown that one can construct an
entanglement monotone (the `Schmidt measure') as the convex-roof
of the logarithm of this quantity \cite{Eisert B 0}.

Finding non-trivial local invariants is quite challenging in general
and can require some sophisticated mathematics. However, for pure
states of some dimensions it is possible to use such invariants to
construct a variety of entanglement quantifiers in a similar fashion
to the tangle. These quantifiers are useful for identifying
different types of multiparty entanglement. We refer the reader to
articles \cite{Miyake,Levay,braunschweig} and references therein for
further details.

\section{Summary, Conclusions, and Open Problems} Quantum entanglement is a rich
field of research. In recent years considerable effort has been
expended on the characterization, manipulation and quantification of
entanglement. The results and techniques that have been obtained in
this research are now being applied not only to the quantification
of entanglement in experiments but also, for example, for the
assessment of the role of entanglement in quantum many body systems
and lattice field theories. In this article we have surveyed many
results from entanglement theory with an emphasis on the
quantification of entanglement and basic theoretical tools and
concepts. Proofs have been omitted but useful results and formulae
have been provided in the hope that they prove useful for
researchers in the quantum information community and beyond. It is
the hope that this article will be useful for future research in
quantum information processing, entanglement theory and its
implications for other areas such as statistical physics.

Despite the tremendous progress in the characterisation of
entanglement in recent years, there are still several major open
questions that remain. Some significant open problems include:

\medskip

{\em Multiparty entanglement:} The general characterisation of
multiparty entanglement is a major open problem, and yet it is
particularly significant for the study of quantum computation and
the links between quantum information and many-body physics.
Particular unresolved questions include:

\begin{itemize} \item {\em Finiteness of MREGS for three qubit states --}
In an attempt to achieve a notion of reversibility in the
multi-partite setting, the concept of MREGS was introduced
\cite{Bennett MREGS}. This was a set of N-partite states for fixed
local dimension from which all other such states may be obtained
asymptotically reversibly. It was hoped for that such a set may
contain only a finite number of states. However, there are
suggestions \cite{Linden PSW 99,Acin multiparty,Galvao PV 00,Wu Z
00} that this may not the case.

\item {\em Distillation results for specific target states --}
In the bi-partite setting the uniqueness of maximally entangled
states led to clear definitions for the distillable entanglement. As
outlined above this is not so in the multi-party setting. Given a
specific interesting multiparty target state (e.g. GHZ states,
cluster states etc.), or set of multiparty target states, what are
the best possible distillation protocols that we can construct? Are
there good bounds that can be derived using multiparty entanglement
measures? Some specific examples have been considered \cite{Murao
PPVK 98,Aschauer DB 04,Goyal CR 06} but more general results are
still missing.
\end{itemize}

{\em Additivity questions:} Of all additivity problems, deciding
whether the entanglement of formation $E_F$ is additive is perhaps
the most important unresolved question. If $E_F$ is additive this
would greatly simplify the evaluation of the entanglement cost. It
would furthermore imply the additivity of the classical capacity of
a quantum channel \cite{Shor,Pom,Audenaert B 04}. Related to the
additivity question is the question of the monotonicity of the
entanglement cost under general LOCC. This may be proven reasonably
straightforwardly if the entanglement cost itself is fully additive.
However, without this assumption no proof is known to the authors,
and in fact a recent argument seems to show that full additivity of
the entanglement cost is equivalent to its monotonicity
\cite{Horodecki private 06}. In addition to $E_F$, there are many
other measures for which additivity is unknown. Examples include the
Distillable Entanglement and the Distillable Key.

\medskip

{\em Distillable entanglement --} Distillable entanglement is a well
motivated entanglement measure of significant importance. Its
computation is however supremely difficult in general and even the
determination of the distillability of a state is difficult. Indeed,
good techniques or algorithms for deciding whether a bipartite state
is distillable or not, and for bounding the distillable
entanglement, are still largely missing.
\begin{itemize}
\item {\em Are there NPT bound entangled states? --} In the bi-partite
setting there are currently three known distinct classes of states
in terms of their entanglement properties under LOCC. These are
the separable states, the non-separable states with positive
partial transpose (which are also non-distillable), and finally
the distillable states. Some evidence exists that there is another
class of states that do not possess a positive partial transpose
but are nevertheless non-distillable \cite{DiVincenzo SSTT 99,Dur
CLB 00}.

\item {\em Bounds on the Distillable entanglement.} Any entanglement
measure provides an upper bound on the distillable entanglement.
Various bounds have been provided such as the squashed
entanglement \cite{Christandl W 03,Christandl 06}, the Rains bound
\cite{Rains 01} and asymptotic relative entropy of entanglement
\cite{Vedral PRK 97,Vedral P 98}. The last two of these coincide
for Werner states \cite{Audenaert EJPVD 01} and it is an open
question whether they always coincide, and whether they are larger
or smaller than the squashed entanglement.
\end{itemize}

{\em Entanglement Measures --} The present article has presented a
host of entanglement measures. Many of their properties are known
but crucial issues remain to be resolved. Amongst these are the
following.

\begin{itemize}
\item {\em Operational interpretation of the relative entropy of
entanglement --} While the entanglement cost and the distillable
entanglement possess evident operational interpretations no such
clear interpretation is known for the relative entropy of
entanglement. A possible interpretation in terms of the
distillation of local information has been conjectured and
partially proven in \cite{Horodecki HHOSSS 04}.

\item {\em Calculation of various entanglement measures --} There
are very few measures of entanglement that can be computed exactly
and possess or are expected to possess an operational
interpretation. A notable exception is the entanglement of
formation for which a formula exists for the two qubit case
\cite{Wootters 98}. Is it possible to compute, or at least derive
better bounds, for the other variational entanglement measures?
One interesting possibility is the 2-qubit case - in analogy to
$E_F$, is there a closed form for the relative entropy of
entanglement or the squashed entanglement?

\item {\em Squashed entanglement --} As an additive, convex, and
asymptotically continuous entanglement monotone the Squashed
entanglement is known to possess almost all potentially desirable
properties as an entanglement measure. Nevertheless, there are a
number of open interesting questions - in particular: (1) is the
Squashed entanglement strictly non-zero on inseparable states, and
(2) can the Squashed entanglement be formulated as a finite
dimensional optimisation problem (with Eve's system of bounded
dimension)?

\item {\em Asymptotic continuity and Lockability questions --} It is
unknown whether measures such as the Distillable Key, the
Distillable Entanglement, and the Entanglement cost are
asymptotically continuous, and it is unknown whether the Distillable
entanglement or Distillable Key are lockable \cite{DiVincenzo HLST
03,Horodecki HHO 04,SynakRadtke H 05}. This is important to know as
lockability quantifies `continuity under tensor products', and so is
a physically important property - if a system is susceptible to loss
of particles, then any characteristic quantified by a lockable
measure will tend to be very fragile in the presence of such noise.
\end{itemize}

{\em Entanglement Manipulation --} Entanglement can be manipulated
under various sets of operations, including LOCC and PPT
operations. While some understanding of what is possible and
impossible has been obtained, a complete understanding has not
been reached yet.

\begin{itemize}
\item {\em Characterization of entanglement catalysis --} For a
single copy of bi-partite pure state entanglement the LOCC
transformations are fully characterized by the theory of
majorization \cite{Nielsen 99,Vidal 99,Jonathan P 99a}. It was
discovered that there are transformations
$|\phi\rangle\rightarrow|\psi\rangle$ such that its success
probability under LOCC is $p<1$ but for which an entangled state
$|\eta\rangle$ exists such that
$|\phi\rangle|\eta\rangle\rightarrow|\psi\rangle|\eta\rangle$ can
be achieved with certainty under LOCC \cite{Jonathan P 99b}. A
complete characterization for states admitting entanglement
catalysis is currently not known.

\item {\em Other classes of non-global operation. Reversibility
under PPT operations --} It is well established that even in the
asymptotic limit LOCC entanglement transformations of mixed states
are irreversible. However in \cite{Audenaert PE 03} it was shown
that that the antisymmetric Werner state may be reversibly
interconverted into singlet states under PPT operations
\cite{Rains 01}. It is an open question whether this result may be
extended to all Werner states or even to all possible states. In
addition to questions concerning PPT operations, are there other
classed of non-global operation that can be useful? If
reversibility under PPT operations does not hold, do any other
classes of non-global operations exhibit reversibility?

\end{itemize}

More open problems in quantum information science can be found in
the Braunschweig webpage of open problems \cite{braunschweig}. We
hope that this list will stimulate some of the readers of this
article into attacking some of these open problems and perhaps
report solutions, even partial ones.

{\it Acknowledgments ---} We have benefitted greatly from
discussions on this topic with numerous researchers over several
years. In relation to this specific article, we must thank K.
Audenaert, F. Brand{\~a}o, M. Horodecki, O. Rudolph, and A.
Serafini for helping us to clarify a number of issues, as well as
G.O. Myhr, P. Hyllus, and A. Feito-Boirac for careful reading and
helpful suggestions. We would also like to thank M. Christandl, J.
Eisert, J. Oppenheim, A. Winter, and R.F. Werner for sharing with
us their thoughts on open problems. This work is part of the
QIP-IRC (www.qipirc.org) supported by EPSRC (GR/S82176/0), the EU
Integrated Project Qubit Applications (QAP) funded by the IST
directorate as contract no. 015848, the EU Thematic network
QUPRODIS (IST-2001-38877), The Leverhulme Trust, the Royal
Commission for the Exhibition of 1851 and the Royal Society.

\renewcommand\bibname{Bibliography}


\begin{thebibliography}{99}
\bibitem{footnote3} This excludes
for example the hydrogen atom from our considerations as it has a
limit point in the spectrum at the ionization level.
\bibitem{counter} The no-go theorem for Gaussian entanglement
distillation is actually quite subtle, as even defining Gaussian
distillation is non-trivial. Consider for example the version of
the no-go theorem proven in \cite{Giedke C 02} - there
entanglement distillation is defined as a process which allows you
to reach a two-mode squeezed state with arbirarily high squeezing.
Although \cite{Giedke C 02} show that this is not possible, from
an operational perspective one can use Gaussian operations to
improve the quality of the entanglement present. For instance,
consider a bi-partite four-mode state $\rho$ which consists of a
two-mode highly entangled state tensored with a product pure
state. Given two copies of $\rho$, Alice and Bob can merely throw
away the product components and are left with a single four-mode
state that possesses greater utility for quantum information
processing. Hence according to an operational definition of
entanglement distillation (say improvement of teleportation
fidelity), it is possible to distill Gaussian entanglement using
Gaussian operations from Gaussian states of 2 $\times$ 2 modes.
This apparent contradiction arises because there are different
ways in which one can quantify the increase of entanglement of
quantum states - in principle this subtle distinction may also be
an issue in the finite dimensional regime.
\end{thebibliography}
\end{document}